\documentclass[3p,  number, single column]{elsarticle}

\usepackage{graphicx}
\usepackage{epsfig}
\usepackage{cite}
\usepackage{amsmath}
\usepackage{amssymb}
\usepackage{url}
\usepackage{fancyvrb}
\usepackage{setspace}

\usepackage{tabularx}
\usepackage{multirow}
\usepackage{booktabs}
\usepackage{dcolumn}

\usepackage{amsmath,amssymb}
\usepackage{graphicx}
\usepackage{dcolumn}
\usepackage{bm,color}
\usepackage{hyperref}
\usepackage{accents}
\usepackage{amssymb,float}
\usepackage{amsmath}
\usepackage{multirow}
\usepackage{siunitx}
\usepackage{tabularx}
\usepackage{booktabs}
\usepackage{url}
\usepackage{soul}
\usepackage{mathrsfs}

\usepackage{mathtools}
\DeclarePairedDelimiter\bra{\langle}{\rvert}
\DeclarePairedDelimiter\ket{\lvert}{\rangle}
\DeclarePairedDelimiterX\braket[2]{\langle}{\rangle}{#1 \delimsize\vert #2}

\begin{document} 

\begin{frontmatter}

\title{Production of Primordial Gravitational Waves in Teleparallel Gravity}

\author{Geovanny A. Rave-Franco}
\ead{geovanny.rave@ciencias.unam.mx}
\address{Instituto de Ciencias Nucleares, Universidad Nacional Aut\'{o}noma de M\'{e}xico, 
Circuito Exterior C.U., A.P. 70-543, M\'exico D.F. 04510, M\'{e}xico.}

\author{Celia Escamilla-Rivera}
\ead{celia.escamilla@nucleares.unam.mx}
\address{Instituto de Ciencias Nucleares, Universidad Nacional Aut\'{o}noma de M\'{e}xico, 
Circuito Exterior C.U., A.P. 70-543, M\'exico D.F. 04510, M\'{e}xico.}

\author{Jackson Levi Said}
\ead{jackson.said@um.edu.mt}
\address{Institute of Space Sciences and Astronomy, University of Malta, Malta, MSD 2080. Department of Physics, University of Malta, Malta, MSD 2080.}


\begin{abstract}
We study the production of primordial gravitational waves in the context of extended teleparallel gravity models and compare them with those of general relativity. Teleparallel gravity has been widely studied in the context of the late universe but not much in the early universe.  Two sources of primordial gravitational waves are considered, vacuum fluctuations and tensor anisotropies within two inflation-compatible backgrounds: a perfect de Sitter and a quasi de Sitter background. We find that in the vacuum case with a perfect de Sitter background, the gravitational waves propagation equation is the same as that of general relativity, however, if the background is promoted to a quasi de Sitter background, the propagation equations are different resulting in an important difference on the tensor spectral index. When tensor anisotropies are included, we compute the most general solution for gravitational waves in terms of a retarded Green’s function and analyze the contributions to the power spectrum from these anisotropies. Finally, we investigate the energy density of these gravitational waves.
\end{abstract}



\end{frontmatter}

\section{Introduction}

General Relativity (GR) has been a successful theory to explain different cosmological phenomena at different cosmological scales in the light of current observational data within the framework of the flat $\Lambda$CDM concordance model \citep{Clifton:2011jh,Planck2018}, which can explain the cosmological dynamics at different stages, particularly, the formation of large-scale structure \citep{Troster2020}, thermal history \citep{Dodelson2021-mi}, the current accelerated expansion \citep{Sahni2004}, and a wide range of other cosmological observations \citep{Riess1998,Bull2016,Hajkarim2020}. However, despite its success, this model faces several theoretical and observational issues, namely, the fine-tuning problem \citep{Perivolaropoulos2022,Martin2012}, coincidence problem \citep{Velten2014,Weinberg1989}, and currently, 
a growing tension between direct and indirect measurements of some prominent cosmological parameters
\citep{DiValentino2021,Abdalla2022}.

Several approaches to solve these problems have been made, some of them include scalar quintessence models \citep{Tin2020,Chimento2003}, high dimensional theories \citep{Kang2020}, a variable cosmological constant \citep{Feng2014}, and extensions of gravity \citep{Bisabr2010,Rudra2015} 
as well as a reexamination of the cosmological principle \citep{Krishnan:2021jmh,Krishnan:2021dyb}, possible early Universe dark energy \citep{Poulin:2023lkg}, and extra degrees of freedom for neutrino species in the early Universe \citep{DiValentino:2021imh,DiValentino:2021rjj}.

In the direction of the latter idea, Teleparallel Gravity (TG) is a gauge theory of gravity where the gravitational field is mediated by torsion instead of curvature \citep{Aldrovandi:2013wha}. Here, a teleparallel equivalent to general relativity (TEGR) can be defined in which all classical phenomena are equivalently predicted, making them dynamically equivalent at this level \citep{Bahamonde:2021gfp,Krssak:2018ywd,Cai:2015emx}. Naturally, this can be extended in similar ways that we do in the regular curvature-based setting. The look for $f(T)$ and $f(T,B)$ extended models in TG have been done to address these issues in the context of the late universe, showing promising in explaining the late time acceleration as a consequence of the geometry of each model \citep{Kadam2022,Franco2020,Mirza2017}, alleviating to some extent the $H_0$ tension \citep{RaveFranco2021,Briffa2022}, the structure formation and inhomogeneities \citep{Nunes2018,Njera2022}, among other topics (see Ref.~\citep{Bahamonde:2021gfp}). 

On the other hand, there have been a few studies made in the context of the early universe, some of them include the study of primordial black holes \citep{ElBourakadi2022}, constraints from inflation \citep{Sahlu2020} and the study of inflation in the context of TG \citep{Raatikainen2019,Chakrabortty2021}. An important effect of the early universe is primordial gravitational waves (GW) since they represent a window to explore the physics of the inflationary era and the first stages of the universe \citep{Ricciardone2017}. The problem of primordial GW in TG was partially addressed in \citep{ElBourakadi2022} but no investigation on the power spectrum 
was performed in this research.

In this work, we will give a first look into the generation of GW as linear tensor perturbations around a flat FLRW background, in the early universe, that is primordial GW. Particularly, we will obtain the solution in TG for GW coming from vacuum fluctuations and the observational implications on the power spectrum and the tensor spectral index. We will also derive the general solution for the GW including an arbitrary possible source of tensor anisotropic stress and see how the power spectrum is changed by the presence of extra geometry given by the non-linearity of the extended $f(T)$ and $f(T,B)$ models, in comparison with GR. As a further step, we will discuss the observational and theoretical implications of GW energy density.

This work is divided as follows: In Sec.~\ref{sec:basics} we will discuss the basic elements of TG, its geometrical setup, and the background cosmology on both $f(T)$ and $f(T,B)$ models. In Sec.~\ref{sec:perturbations} we discuss the linear perturbation theory in TG  from a transverse and traceless tensor perturbation, we will study the production of GW coming from vacuum fluctuations and tensor anisotropies. Finally, in Sec.~\ref{sec:discussions} we discuss the principal results of this work and its observational implications, and in Sec.~\ref{sec:conclusions} we present our conclusions. In this work, Greek indices $\mu,\nu, \ldots$ refer to spacetime indices, and capital Latin indices $A,B, \ldots$ refer to a tangent or Minkowski spacetime indices.


\section{Basics of teleparallel gravity and its cosmology}
\label{sec:basics}

TG is a gauge theory of translations invariant under local Lorentz transformations \citep{Aldrovandi:2013wha}. The geometrical setup of TG consists of two fundamental objects \citep{Hohmann2019}, the tetrad fields $\boldsymbol{e} \in \Omega^1(\mathcal{M},\mathbb{R}^{1,3})$, a set differential $1$-forms on $M$ assuming values on the Minkowski space $\boldsymbol{e} = \{e^A\}^{3}_{A=0}$, and the spin connection  $\boldsymbol{\omega} \in \Omega^1(\mathcal{M},\mathfrak{so}(1,3))$ a  differential $1$-form assuming values on the Lie Algebra of the Lorentz Group. On a local coordinate system $U \subset \mathcal{M}$ of the base manifold $\mathcal{M}$, these  objects read as
\begin{equation}
    e^A_x = e^A_{\ \mu}(x)dx^{\mu}\,, \quad \boldsymbol{\omega}  = \frac{1}{2}\omega^{AB}_{\ \ \mu}dx^{\mu}\otimes S^{AB} \in \Gamma\left(T^*U \otimes \mathfrak{so}(1,3) \right)\,, \quad x \mapsto (x,\boldsymbol{\omega}_x)\,,
\end{equation}
where $S^{AB}$ are the generators of the Lorentz group, $\mathfrak{so}(1,3)$ the Lie Algebra of the Lorentz group, $\Gamma(T^*U)$ the sections over the cotangent bundle and  $\boldsymbol{\omega}_x = \frac{1}{2}\omega^{AB}_{\ \ \mu}(x)dx^{\mu}\otimes S^{AB} \in T^*U \otimes \mathfrak{so}(1,3)$ \citep{Socolovsky2012a,Socolovsky2012b}. However, in TG the tetrad fields and the spin connection are not any arbitrary fields satisfying the aforementioned definitions but particular fields as we shall see later on in this section. The tetrad fields $e^A =  e^A_{\ \mu}dx^{\mu}$ and $E_{A} = E_{A}^{\ \mu}\partial_{\mu}$ constitute the non-coordinate basis of the sections over the tangent bundle and cotangent bundles  $\Gamma(T U)$ and $\Gamma(T^*U)$ in every local coordinate system $U$ of the manifold $\mathcal{M}$ such that 
\begin{equation}
    e^{A}_{\ \mu}E_{A}^{\ \nu}= \delta_{\mu}^{\nu}\,, \quad \text{and} \quad e^{A}_{\ \mu}E_{B}^{\ \mu}= \delta_{A}^{B}\,,
\end{equation}
and at every point of the base manifold $p \in \mathcal{M}$ they diagonalize the metric \citep{Blau2011}
\begin{equation}
    g_{AB}(p)=\eta_{AB}(p) = g_{\mu \nu}(p)E^{\mu}_{A}(p)E^{\nu}_B(p)\,, \label{diagonalization}
\end{equation}
or analogously,
\begin{equation}\label{tetradandmetric}
    g_{\mu \nu}(p) = e^{A}_{\ \mu}(p)e^{B}_{\ \nu}(p)\eta_{AB}(p)\,.
\end{equation}
Under diffeomorphism over the base manifold $\mathcal{M}$ and local Lorentz transformation $\Lambda^{A}_{\ B}(\boldsymbol{x}) \in SO^+(1,3)$ over the Minkowski space, the tetrad fields transform as
\begin{equation}
    e^{A}_{\ \mu '} = \frac{\partial x^{\nu}}{\partial x^{\mu '}}e^{A}_{\ \nu }\,, \quad \text{and} \quad e^{A}_{\ \mu} = \Lambda ^{A}_{\ B}e ^{B}_{\ \mu}\,.
\end{equation}
In the case of the spin connection, we aim to describe a globally flat manifold but with non-trivial geometry given by the torsion, hence, we require the curvature $2$-form to satisfy
\begin{equation}\label{curvature 2-form}
    \mathbf{R} = \frac{1}{4}R^A_{\text{   }B \nu \mu}S_A^B dx^{\nu}\wedge dx^{\mu} \quad \text{with} \quad R^A_{\text{   }B\nu\mu} = \partial_{\nu}\omega^{A}_{\text{   }B\mu} - \partial_{\mu}\omega^{A}_{\text{   }B\nu} + \omega^A_{ \text{   }D \nu}\omega^D_{\text{   }B \mu} - \omega^A_{\text{   }D \mu}\omega^D_{\text{   }B \nu}\equiv 0\,,
\end{equation}
which is achieved by 
\begin{equation}\label{purelyinertialspinconnection}
    \omega^{A}_{\ B \mu} = \Lambda^{A}_{\ C}(x)\partial_{\mu}\Lambda^{C}_{\ B}(x)\,,
\end{equation}
called \textit{the purely inertial spin connection}, accounting for the inertial effects of the theory. The purely inertial spin connection is the portrayal of a zero spin connection in an arbitrary Lorentz frame
\begin{equation}
    {\omega^{A}_{\ B\mu}}(\boldsymbol x) = \Lambda^{A}_{\ C}(\boldsymbol x){\omega^{C}_{\ D \mu}}'(\boldsymbol x)\Lambda_{B}^{\ D}(\boldsymbol x) + \Lambda^A_{\ C}(\boldsymbol x)\partial_{\mu} \Lambda^C_{\ B}(\boldsymbol x) \quad \text{with} \quad {\omega^{C}_{\ D \mu}}' = 0\,,
\end{equation}
and thus can be transformed into a zero spin connection using the inverse Lorentz transformation \citep{Hohmann2019}. The process of choosing a vanishing spin connection ${\omega^{A}_{\ B\mu}} =0$ is called \textit{the Weitzenböck gauge}, and can only be applied when the spin connection field equations are identically satisfied. 
Finally, TG requires a pair of tetrad and spin connections where the torsion $ 2$ form satisfies
\begin{equation}\label{torsion 2-form}
    \mathbf{T} = \frac{1}{2}T^{A}_{\nu \mu}P_A dx^{\nu} \wedge dx^{\mu}\,, \quad \text{with} \quad  T^A_{\text{   } \nu \mu} = \partial_{\nu} e^A_{\mu} - \partial_{\mu}e^A_{\nu} + \omega^A_{\text{   }C \nu}e^{C}_{\mu} - \omega^A_{\text{   }C \mu}e^C_{\nu} \neq 0\,,
\end{equation}
where $P_A$ are the generators of the translational group. Hence, TG is a gravitational theory where gravity is a manifestation of the torsion and not of curvature of spacetime. The relation between the torsion and curvature tensors with the torsion and curvature $2$-forms are given by
\begin{equation}
    R^{\rho}_{\ \gamma \nu \mu} = E^{\rho}_A e^{B}_{\gamma}R^A_{\ B \nu \mu}, \quad T^{\rho}_{\ \nu \mu}=E_{A}^{\ \rho}T^{A}_{\ \nu \mu} = -2\Gamma^{\rho}_{\left[ \nu \mu \right]}\,,\label{relation tensors and 2-forms}
\end{equation}
where the TG connection is given by 
\begin{equation}\label{LinearConnectionFromSpin}
    \Gamma_{\nu \mu}^{\rho} \equiv E_{A}^{\rho}\partial_{\mu}e^{A}_{\nu} + E_{A}^{\rho}\omega^{A}_{\text{   }B \mu}e^{B}_{\nu} = E_{A}^{\rho}\mathscr{D}_{\mu}e^{A}_{\nu}\,.
\end{equation}
Throughout this work, we will work on the Weitzenböck gauge, since for the case of cosmological symmetry, the field equations associated to variations with respect to the spin connection are satisfied \citep{Hohmann2019}.

Under this scheme, we can write the TG scalars\footnote{We can extend this to any \textit{object}, for instance, the Riemann tensor can be written w.r.t the Levi-Civita connection using the contorsion tensor.}
in terms of the Levi-Civita connection with the use of the \textit{contorsion tensor}\footnote{All geometric objects with an over circle are computed using the Levi-Civita connection.}
\begin{equation}
    \Gamma^{\rho}_{\mu \nu} = \accentset{\circ}{\Gamma} ^{\rho}_{\mu \nu} + K^{\rho}_{\mu \nu}\,.
\end{equation}
Particularly, if we compute the Ricci scalar, we arrive at \citep{Golovnev2017}
\begin{equation}\label{ricciscalar}
    R = \accentset{\circ}{R} + T - B\,,
\end{equation}
with 
\begin{align}
    T &=T^{\alpha}_{\text{   }\sigma \rho}S_{\alpha}^{\text{   }\sigma \rho}= \frac{1}{4}T^{\mu \nu \lambda}T_{\mu \nu \lambda} + \frac{1}{2}T^{\mu \nu \lambda}T_{\nu \mu \lambda} - T^{\mu}T_{\mu}\,, \label{torsionscalar} \\
    S_{\alpha}^{\text{   }\sigma \rho} &= \frac{1}{4}(T_{\alpha}^{\text{   }\sigma \rho} + T^{\rho\sigma}_{\hspace{0.3cm} \alpha} - T^{\sigma \rho}_{\hspace*{0.3cm}\alpha} - 2 T^{\lambda\sigma}_{\hspace*{0.3cm}\lambda}\delta_{\alpha}^{\rho} + 2 T^{\lambda \rho}_{\hspace*{0.3cm}\lambda}\delta_{\alpha}^{\sigma}) = \frac{1}{2}\left( K_{\hspace{0.3cm} \alpha}^{\sigma \rho} +  T^{\sigma}\delta_{\alpha}^{\rho} -  T^{\rho}\delta_{\alpha}^{\sigma} \right)\,, \label{superpotentialtensor} \\
    B &= \frac{2}{e}\partial_{\mu}\left( e T^{\mu} \right) = 2\nabla_{\mu}T^{\mu}\,. \label{boundaryterm}
\end{align}
In TG, we arrive at the following equation
\begin{equation}
     \accentset{\circ}{R} =  - T + B\,,
\end{equation}
which is arrived at due to the vanishing of curvature in this setting. Through this expression and substituting the tetrad determinant $\sqrt{-g}=e$ into the Einstein-Hilbert action \citep{Carroll2004}, considering that $B$ is a boundary term as a vanishing object at linear level, we obtain the action of the TEGR
\begin{equation}\label{TEGRaction}
    S_{\text{TEGR}} = -\frac{1}{2\kappa} \int eT d^4x + \int e\mathcal{L}_{\text{m}}d^4x\,,
\end{equation}
with $\kappa = 8\pi G$. The field equations of the TEGR action (\ref{TEGRaction}) recover the same equations as in GR at the level of the equations of motion for particular systems, leaving us with the same phenomenology of GR. In the same spirit as $f(\accentset{\circ}{R})$ \citep{Sotiriou:2008rp,Faraoni:2008mf,Capozziello:2018qcp} extensions of TEGR have been gaining interest in recent years \citep{Farrugia:2020fcu,Farrugia:2018gyz,Bahamonde:2015zma}, particularly, we look for extensions of TEGR in the form of an arbitrary functional $f(T,B)$ into the action
\begin{equation}
    S_{f(T,B)} = -\frac{1}{2\kappa}\int ef(T,B)d^4x + \int e\mathcal{L}_{\text{m}}d^4x\,,
\end{equation}
such that variations concerning the tetrad gives the following field equations
\begin{align}\label{f(T,B) field equations}
    W^{\lambda}_{\ \nu} \equiv & \delta^{\lambda}_{\nu}\accentset{\circ}{\square}f_B -  \accentset{\circ}{\nabla}^{\lambda}\accentset{\circ}{\nabla}_{\nu}f_B + \frac{1}{2}f_BB\delta^{\lambda}_{\nu} + 2\left[\partial_{\mu}f_B + \partial_{\mu}f_T \right]S_{\nu}^{\text{   }\mu \lambda} + \frac{2}{e}e^A_{\nu}f_T\partial_{\mu}(e S_A^{\text{   }\mu \lambda}) \\ \notag & - 2f_TT^{\alpha}_{\text{   }\mu \nu}S_{\alpha}^{\text{   }\lambda \mu} - \frac{1}{2}f\delta^{\lambda}_{\nu} =  \kappa \Theta^{\lambda}_{\ \nu}\,,
\end{align}
where the energy-momentum tensor is defined as
\begin{equation}\label{Energy-momentum tensor TG}
    \Theta^{\lambda}_{\ \nu} = e^{A}_{\ \nu}\theta^{\lambda}_{\ A}, \quad \theta_{\ A}^{\lambda} = \frac{1}{e} \frac{\delta (e \mathcal{L}_{\text{m}})}{\delta e^A_{\ \lambda}}\,.
\end{equation}
If the function is not dependent on the boundary term $f(T,B)=f(T)$, we recover the field equations of $f(T)$ gravity, and if $f(T,B)=f(T)=-T$ we recover the TEGR field equations \citep{Bahamonde:2021gfp}. The spin connection field equations turn out to be written as the anti-symmetry of these equations, given by
\begin{equation}
    W_{[\mu\nu]} = 0\,,
\end{equation}
where square brackets denote the antisymmetry operator.

In the case of cosmological symmetry, where the spacetime manifold is foliated with maximally symmetric space-like slices $\mathcal{M}=\mathbb{R}\times \Sigma$ \citep{Carroll2004,Wald} showing isotropy and homogeneity, we work with the FLRW metric given by 
\begin{equation}
     ds^2 = dt^2 - a(t)^2\left( dx^2 +dy^2 +dz^2 \right) ,\label{FLRW metric}
\end{equation}
in the spatially flat case $k=0$, with Cartesian coordinates
The corresponding tetrad field which exhibits the cosmological symmetry in the Weitzenböck gauge is given by
\begin{equation}\label{BackgroundTetrad}
    e^A_{\ \mu} = \text{diag}(1,a(t),a(t),a(t))\,,
\end{equation}
which turns out to be compatible with the Weitzenb\"{o}ck gauge. When working with cosmological symmetry, the antisymmetric part of the field equations is completely satisfied \citep{Hohmann2019}, leaving us the task to solve only the symmetric part of the field equations. Using the background tetrad (\ref{BackgroundTetrad}), we compute the torsion and superpotential tensors to be \citep{Bahamonde2021}
\begin{equation}
    T^{i}_{\ 0 j} = H\delta^i_j\,, \quad S_{i}^{\ 0j} = -H\delta^i_j\,,
\end{equation}
with the torsion scalar and boundary terms are 
\begin{equation}
    T = -6H^2\,, \quad \text{and} \quad B=-6(3H^2 + \dot{H})\,,
\end{equation}
and $H = \dot a/a$ is the Hubble parameter in cosmic time. The background field equations for $f(T,B)$ are then given by
\begin{align}
    3H\dot{f}_B - 3H^2(3f_B + 2f_T) - 3f_B\dot{H} - \frac{1}{2}f(T,B) = \kappa \rho\,, \label{f(T,B) first friedmann} \\
    -(3H^2 + \dot{H})(2f_T + 3f_B) - 2H\dot{f}_T + \ddot{f}_B - \frac{1}{2}f(T,B)  = - \kappa P\,.
\end{align}
The energy density $\rho$ and pressure $P$ correspond to those of a perfect fluid.

From now on, we will focus only on these latter equations

for simplicity of the discussion, taking into account that $f(T,B)=f(T)$ and $f(T,B)=-T$ recover the appropriate equations of each theory. We have been working so far with cosmic time, however, in the case of linear perturbation theory, we will switch to conformal time \citep{Piattella2018} 
\begin{equation}
    ad\eta = dt \quad  \rightarrow  \quad \eta - \eta_i = \int_{t_i}^{t} \frac{dt'}{a(t')}\,,\label{eq:conformal}
\end{equation}
where the background tetrad becomes
\begin{equation}
    e^{A}_{\ \mu} = a(\eta)\text{diag}(1,1,1,1)\,,
\end{equation}
and the primes will indicate derivatives concerning conformal time.


\section{Tensor perturbations in TG}
\label{sec:perturbations}

In this section, we explore the tensor perturbations around a flat FLRW background in order to probe the production of primordial GWs in TG. We break the background cosmological symmetry by considering the first order deviation of the background tetrad\footnote{The quantities with an overbar are background quantities, which in this context exhibit the cosmological symmetry.}
\begin{equation}
    e^{A}_{\ \mu} = \bar e^{A}_{\ \mu} +  \delta e^{A}_{\ \mu}\,,
\end{equation}
such that $| \delta e^{A}_{\ \mu}|\ll 1$, such that the corresponding metric perturbation is given by \citep{Hohmann2021}
\begin{equation} \label{pmetricfromtetrtad}
    \delta g_{\mu \nu} = 2 \tau_{(\mu \nu)} \quad \text{where} \quad \tau_{\mu \nu}= \bar e^{A}_{\ \mu}\delta e^{B}_{\ \nu}\eta_{AB}\,.
\end{equation}
We consider the linear perturbation to be a symmetric transverse and traceless tensor perturbation, in conformal time, given by 
\begin{equation}\label{ptetradtensor}
    \delta e^{A}_{\ \mu} = \frac{a(\eta)}{2}
    \begin{pmatrix}
        0 & 0 \\
        0 & h_{ij} \\
    \end{pmatrix}\,,
\end{equation}
such that the linear perturbation of the torsion and superpotential tensors are given by
\begin{equation}
    \delta T^{i}_{\ 0 j} = \frac{1}{2}h_{ij}', \quad \delta T^{i}_{\ jk} = \frac{1}{2}(\partial_j h_{ik} - \partial_{k}h_{ij}), \delta S_{0}^{\ 0i}=0, \quad \delta S_{i}^{\ 0j} = \frac{1}{4a^{2}}h_{ij}', \quad \delta S_{i}^{\ jk} = -\frac{1}{4a^2}(\partial_{j}h_{ik} - \partial_k h_{ij})\,.
\end{equation}
The torsion scalar and boundary terms are given by
\begin{equation}
    \delta T = \delta B = 0\,,
\end{equation}
from where we observe the Lagrangian functional form to have dependencies
\begin{equation}
    f(T,B) = f(\bar T, \bar B)\,,
\end{equation}
that is, they only depend on the background torsion scalar and boundary term. Thus, it turns out that every time the function or its derivatives appear in calculations, we should keep in mind that they are dependent only on background quantities. Finally, the perturbed part on the left-hand side of the field equations is given by
\begin{equation}
    \delta W^{i}_{\ j} = \frac{f_T}{2a^2} \delta ^{ik}\left(h''_{kj} + [2 + \nu]\mathcal{H}h'_{kj} - \nabla^2 h_{kj} \right)\,, \label{perturbedGW}
\end{equation}
from where, including tensor anisotropic effects, we arrive at the following equation in Fourier space 
\begin{equation}\label{GWTG}
    h''_{kj} + [2 + \nu]\mathcal{H}h'_{kj}  + k^2 h_{kj} = \frac{16\pi G a^{2}}{f_T}\pi_{ij}^T\,,
\end{equation}
where $\nu = \frac{1}{\mathcal{H}}\frac{f'_T}{f_T}$ and $\mathcal{H}= a'/a$ \citep{Chen:2010va,Bahamonde:2022ohm}
Eq. (\ref{GWTG}) is the propagation equation for a massless GW that propagates at the speed of light $c$, with Planck mass run rate $\nu$, and which agrees with the literature form of this evolution equation \citep{Nunes:2018evm}. Notice that since we are working with a symmetric transverse and traceless tensor perturbation, only two degrees of freedom are propagating, namely, the so-called $+$ and $\times$ polarizations. We now proceed to investigate the production of these GW coming from vacuum fluctuations and possible tensor anisotropies. We will study the effects of the extensions on the power spectrum and the energy density of GW. In Refs.~\citep{Bahamonde2021,Bahamonde:2021gfp}, the condition $f_T <0$ is required for stability of the tensor perturbation.

We begin by considering an inflationary scenario, an exponentially accelerated expansion in the early universe driven by a scalar field that solves several theoretical problems on the $\Lambda$CDM model \citep{Ryden2017}. By quantizing the GW field as
\begin{equation}\label{GWquntize}
	h_{ij}(\eta,\mathbf x) = \sum_{\lambda=\pm 2}\int\frac{d^3\mathbf k}{(2\pi)^3}\left[h(\eta,k)e^{i\mathbf k\cdot\mathbf x}a(\mathbf k,\lambda)e^{\lambda}_{ij}(\hat{k})+ h^*(\eta,k)e^{-i\mathbf k\cdot\mathbf x}a^{\dagger}(\mathbf k,\lambda){e^{\lambda}}^*_{ij}(\hat{k})\right]\;\,,
\end{equation}
with $e_{ij}^{\lambda}$ the polarization tensors, and we have defined $h^{\lambda}(\eta,\mathbf{k}) = h(\eta,k)a(\mathbf k,\lambda)$ such that \citep{Klose2022b}
\begin{equation}
    h_{ij} = \sum_{\lambda}e_{ij}^{\lambda}h^{\lambda}, \quad \sum_{i,j}e_{ij}^{\lambda}{e_{ij}^{\lambda '}}^*=\delta^{\lambda \lambda'}\,. \label{Orthonormality}
\end{equation}
The operators $a(\boldsymbol k,\lambda),a^{\dagger}(\boldsymbol k,\lambda)$ are quantum operators that annihilate and create gravitons of momentun $k$ and helicity $\lambda$, and satisfy the canonical commutation relations \citep{Piattella2018}
\begin{equation}\label{commutationrelationsGW}
	\left[a(\mathbf k,\lambda), a(\mathbf k',\lambda')\right] = 0\;\,, \qquad \left[a(\mathbf k,\lambda), a^\dagger(\mathbf k',\lambda')\right] = (2\pi)^3\delta^{(3)}(\mathbf k - \mathbf k')\delta_{\lambda\lambda'}\;\,.
\end{equation}
We will assume the state of the universe to be in a vacuum quantum state $\ket{0}$ such that
\begin{equation}
    a(\boldsymbol k,\lambda)\ket{0}=0\,,
\end{equation}
such that in the infinite past $\eta \to -\infty$, the modes $h(\eta,k)$ resemble the quantisation of a free field in Minkowski space \citep{Piattella2018}\citep{Kundu2012}. This state is called the Bunch-Davies vacuum state.

The correlator of the quantized tensor field is given by
\begin{equation}
	\langle 0|h_{ij}(\eta,\mathbf x)h_{lm}(\eta,\mathbf x')|0\rangle = \int\frac{d^3\mathbf k}{(2\pi)^3}|h(\eta,k)|^2e^{i\mathbf k\cdot(\mathbf x - \mathbf x')}\Pi_{ij,lm}(\hat{k})\;\,, \label{correlatorGW}
\end{equation}
where the sum over helicities is given by \citep{weinberg2008cosmology}
\begin{align}
	\Pi_{ij,lm}(\hat k) & \equiv \sum_{\lambda = \pm 2}e_{ij}(\hat{k},\lambda)e^*_{lm}(\hat{k},\lambda)\;\\
    & = \delta_{im}\delta_{jn} + \delta_{in}\delta_{jm} - \delta_{ij}\delta_{mn} + \delta_{ij}\hat{k}_m\hat{k}_n + \delta_{mn}\hat{k}_i\hat{k}_j - \delta_{im}\hat{k}_j\hat{k}_n - \delta_{jn}\hat{k}_i\hat{k}_m - \delta_{in}\hat{k}_j\hat{k}_m - \delta_{jm}\hat{k}_i\hat{k}_n\nonumber\\
	& + \hat{k}_i\hat{k}_j\hat{k}_m\hat{k}_n\;\,.
\end{align}
In superhorizon limits $k|\eta|\to 0$, particularly when the $k$-mode re-enters the horizon, the quantum field can be seen as a classical field, hence, we expect the correlator to become a classical average ensemble, from where assuming gaussianity, we identify the power spectrum to be
\begin{equation}
    P_h(\eta,k) = |h(\eta,k)|^2\,,
\end{equation}
and the dimensionless power spectrum \citep{Baumann2022-sw}
\begin{equation}
    \Delta_h^2(\eta,k) = \frac{k^3}{2\pi^2}|h(\eta,k)|^2\,.
\end{equation}
Now, we proceed to analyze the power spectrum for GW in TG coming from vacuum fluctuations and also the contributions from tensor anisotropies.


\subsection{Power spectrum from vacuum fluctuations}
\label{subsec:PS2}

Vacuum fluctuations are generated where the inflaton field dominates. Since the inflaton is a scalar field, no anisotropic stress is present during this epoch \citep{weinberg2008cosmology}, thus, the propagation equation for GW in TG is given by
\begin{equation}
     h_{ij}'' + [2 + \nu]\mathcal{H}h_{ij}' + k^2h_{ij} = 0\,.
\end{equation}
If we multiply by the polarization tensor and sum over helicities, we arrive at
\begin{equation}
    {h^{\lambda}}'' + [2 + \nu]\mathcal{H}{h^{\lambda}}' +  k^2 h^{\lambda}=0\,,
\end{equation}
from which we observe that both polarization states satisfy the same propagation equation in a vacuum, allowing us to work with $h^{\lambda}\equiv h$ to simplify notation. We will now consider two cases of inflation, one with a perfect de Sitter expansion and another with a quasi de Sitter expansion modeled by the slow-roll parameter $\epsilon$.

From now on, we will be working with canonically normalized tensor perturbations \citep{Piattella2018,Klose2022b}
\begin{equation}
    \hat{h}_{ij} = \frac{h_{ij}}{\sqrt{32\pi G}}  \equiv \frac{M_{\text{pl}}}{2}h_{ij}\,,
\end{equation}
for convenience


\begin{itemize}
\item \textbf{De Sitter background.}
A perfect de-sitter background is described by
\begin{equation}
    H = H_{\Lambda} = \text{const} \rightarrow \dot{H} = 0\,,
\end{equation}
which has important implications for GWs in TG. Recalling that for tensor perturbation the functional is only dependent on background quantities $f(T,B) =  f(\bar T, \bar B)$, we observe that, since $\bar T = -6H^2$ and $\bar B = -6(3H^2 + \dot{H})$, we have
\begin{equation}
    f'_T = \frac{dt}{d\eta}\dot{f}_T = \frac{dt}{d\eta} \left(f_{TT}\dot{\bar T} + f_{TB}\dot{\bar B} \right)= 0\,,
\end{equation}
from where the $\nu$ 
parameter vanishes, thus, the equation becomes
\begin{equation}\label{GREquivalentEquation}
    {\hat h}'' + 2\mathcal{H}{\hat h}' + k^2{\hat h} = 0\,,
\end{equation}
which is the equation for GW in GR \citep{Boyle2008}. Therefore, in a vacuum with a perfect de Sitter background TG and GR produce the same gravitational waves. To solve Eq.~\eqref{GREquivalentEquation}, we need to recast it as an equation with no damping term \citep{Piattella2018}, which is achieved by performing the change of variable
\begin{equation}\label{GRchange}
    \hat{h} = \frac{g}{a}\,,
\end{equation}
from where Eq.~\eqref{GREquivalentEquation} becomes
\begin{equation}\label{EqNoDamping}
    g'' + \left(k^2 - \frac{2}{\eta^2} \right)g = 0\,,
\end{equation}
where we have used the fact that $a(\eta) = -\frac{1}{H_{\Lambda}\eta}$ in a perfect de Sitter background. The solutions of Eq.~\eqref{EqNoDamping} are given by
\begin{equation}\label{VacuumSolGR}
    g \subset \left\lbrace C(k)k\left(1 - \frac{i}{k \eta} \right)e^{-ik\eta}, \quad  C^*(k)k\left(1 + \frac{i}{k \eta} \right)e^{ik\eta}\right\rbrace\,,
\end{equation}
where $C(k)$ is an arbitrary integration function of $k$.
However, to satisfy the initial conditions in the infinite past, we need to choose the first solution
\begin{equation}
    g(\eta,k)= C(k)k\left(1 - \frac{i}{k \eta} \right)e^{-ik\eta}\,,
\end{equation}
and evaluate it in the infinite past  $k|\eta_i|\to \infty$, or $k|\eta_i| \gg 1$, imposing the condition of the field to be a free field
\begin{equation}
    g(\eta_i,k) = C(k)ke^{-ik\eta_i} = \frac{1}{\sqrt{2k}}e^{-ik\eta_i} \rightarrow C(k)k = \frac{1}{\sqrt{2k}}\,,
\end{equation}
and then we can just return to the original $h$ field given by
\begin{equation}
    \hat h = \frac{1}{a}\frac{e^{-i k \eta}}{\sqrt{2k}}\left(1 - \frac{i}{k \eta} \right)\,,
\end{equation}
from where the power spectrum, taking into account the canonical normalization, is
\begin{equation}
    P_h(\eta,k) = \frac{16 \pi G}{k}H^2_{\Lambda}\eta^2 \left( 1 +  \frac{1}{k^2 \eta^2} \right)\,,
\end{equation}
and the dimensionless power spectrum, considering that there are two polarization states \citep{Boyle2008}, is given by 
\begin{equation}
    \Delta^2_h (\eta, k) \equiv \frac{d \bra{0}\hat{h}^2_{ik}\ket{0}}{d \ln k} = 64 \pi G \frac{k^3}{2\pi^2} |\hat h(\eta,k)|^2= \frac{2H^2_{\Lambda}}{\pi^2 M_{\text{pl}}^2}[1 + k^2\eta^2]\,. 
\end{equation}
On the super horizon limit $k|\eta|\to 0$, we recover the scale-invariant dimensionless power spectrum 
\begin{equation}
    \Delta^2_h (k)=\frac{2}{\pi^2}\frac{H^2_{\Lambda}}{ M_{\text{pl}}^2}\,,
\end{equation}
which is the same as GR as stated before \citep{Baumann2022-sw}. As we have seen, in the case of a perfect de Sitter background, there is no difference between the GW of GR and TG, resulting in the same power spectrum. this prompts us to work in a slightly different scenario, a quasi-de Sitter background.
 
\item \textbf{Quasi de Sitter background.}
The quasi de Sitter background is defined by a slowly varying derivative of the Hubble factor
\begin{equation}
    \dot{H} = -\epsilon H^{2}_{\Lambda}\,, \quad a(\eta) = \frac{1}{H_{\Lambda}}\frac{1}{|\eta|^{1+\varepsilon}}\,, \quad \text{where} \quad |\eta|=-\eta\,,
\end{equation} 
with $|\epsilon|\ll 1$ is the first slow roll parameter. Since the slow roll parameter is small, all calculations will be performed in the first order in $\epsilon$. Now, we need to compute the $\nu$ parameter for a general $f(T,B)$ functional. The idea is to Taylor expand at first order in $\epsilon$ the $\nu$ parameter
\begin{equation}
    \nu(\epsilon) = \nu(\epsilon=0) + \epsilon\partial_{\epsilon}\nu|_{\epsilon=0}\,,
\end{equation}
such that $\nu(\epsilon=0)=0$ as we have argued previously since at $\epsilon \to 0$, $f'_T =0$. Using the same argument we obtain
\begin{equation}
    \partial_{\epsilon}\nu|_{\epsilon=0} = -\frac{\eta}{f_{T_{\Lambda}}}\partial_{\epsilon}f'_T|_{\epsilon = 0}\,,
\end{equation}
where introducing the notation  $T_{\Lambda} = -6H_{\Lambda}^2$ and $B_{\Lambda}= 3T_{\Lambda}$, we arrive at
\begin{equation}
    \partial_{\epsilon}\nu|_{\epsilon=0} = 2 \left( \frac{f_{T_{\Lambda}T_{\Lambda}}|T_{\Lambda}| + f_{T_{\Lambda}B_{\Lambda}}|B_{\Lambda}|}{f_{T_{\Lambda}}}\right)\,,
\end{equation}
from where we obtain the $\nu$ parameter to be
\begin{equation}
    \nu = 2\gamma \epsilon\,, \quad \text{with} \quad \gamma = \left( \frac{f_{T_{\Lambda}T_{\Lambda}}|T_{\Lambda}| + f_{T_{\Lambda}B_{\Lambda}}|B_{\Lambda}|}{f_{T_{\Lambda}}}\right)\,.
\end{equation}
Nevertheless, the stability condition $f_T < 0$ needs to be fulfilled and it depends on the particular form of the function. In the literature there have been several models that can achieve these conditions and have been tested against observations \citep{Bahamonde:2016grb,Bahamonde2019,Rezazadeh2016}, we will expose some of these models next \footnote{Recall that $\delta T = \delta B=0$.}. 
\begin{enumerate}
\item \ul{\textit{Power law models}}. There are models of $f(T,B)$ and $f(T)$ in the form of power law as
\begin{equation}
    f( T) = -T + f_0(- T)^m\,, \quad f( T, B) = T + f_0(- T)^m + f_1(-B)^m\,.
\end{equation}
These models produce the same type of GW since both of them imply
\begin{equation}
    f_{T} = -1 - mf_0(- T)^{m-1}\,,
\end{equation}
and then have associated the same $\nu$ parameter. From observations \citep{dosSantos2022,Escamilla-Rivera:2019ulu} we have $m,f_0>0$ such that $f_T <-1$, fulfilling the stability condition.
\item \ul{\textit{Mixed power law model.}} In this case, the function is given by
\begin{equation}
    f(T,B) = -T + f_0 (-T)^{m}(-B)^n\,,
\end{equation}
where the partial derivative w.r.t the torsion scalar is
\begin{equation}
    f_T = -1 - mf_0 (-T)^{m-1}(-B)^n\,,
\end{equation}
and according to observations \citep{Escamilla-Rivera:2019ulu} we have $f_0,m>0$ implying $f_T<-1$, also fulfilling the stability condition.
\item \ul{\textit{Exponential model.}} In this case, we deal with a $f(T)$ model given by
\begin{equation}
    f(T) = -T  + \beta T_{\Lambda}(1 - e^{-q T/T_{\Lambda}})\,.
\end{equation}
In the limit, $q\to \infty$ the model is reduced to be GR \citep{Li2018}. Using this model we have
\begin{equation}
    f_T = -1 + \beta q e^{-q T/T_{\Lambda}}\,,
\end{equation}
where the stability condition $f_T <0$ requires $\beta q <0$ which is fulfilled, since observations requires $q>0$ and $\beta <0$ \citep{Nesseris2013}.
\end{enumerate}
The gamma parameter for these models is given by
\begin{align}\label{gammaparamater}\Large{
    \gamma = \left\lbrace \begin{array}{ll}
    \frac{f_0 m(1-m)|T_{\Lambda}|^{m-1}}{1 +mf_0|T_{\Lambda}|^{m-1}}\,, & \quad \text{\normalsize{for the power law models}}, \\
    \frac{f_0 m(1-m - n)|T_{\Lambda}|^{m-1}|B_{\Lambda}|^n}{1+mf_0|T_{\Lambda}|^{m-1}|B_{\Lambda}|^n}\,, & \quad \text{\normalsize{for the mixed power law model,}} \\
    \frac{\beta  q^2 }{\beta  q-e^q}\,, & \quad \text{\normalsize{for the exponential model.}}
    \end{array} \right.}
\end{align}
With this in mind, we are now in a position to compute the power spectrum for GW coming from vacuum fluctuations. The equation for GW for vacuum fluctuations is 
\begin{equation}
    {\hat h}'' + [2 + \nu]\mathcal{H}{\hat h}' + k^2 {\hat h}= 0\,,
\end{equation}
where we again aim to recast it as the equation of a harmonic oscillator without the damping effect, however, the change of variable in (\ref{GRchange}) is not suitable for that purpose. For an arbitrary $\nu$ parameter, the appropriate change of variable is given by
\begin{equation}
    h(\eta,k) = f(\eta)g(\eta,k), \quad \text{with} \quad  f(\eta) = \frac{1}{a(\eta)}\text{Exp}\left(-\frac{1}{2}\int \mathcal{H}\nu d\eta \right)\,. \label{AnsatzSolution}
\end{equation}
With this change of variable, the equation for GW in TG in the vacuum case reduces to
\begin{equation}
    g'' + g\left(k^2 + \frac{f''}{f} + [2 + \nu]\mathcal{H}\frac{f'}{f} \right) = 0\,, \label{GWinTGnodamping}
\end{equation}
where in the case of a quasi-de Sitter background reads as
\begin{equation}\label{GWinTGnodampingQuasideSitter}
    g'' + g\left(k^2 - \frac{2 + 3(1 + \gamma)\epsilon}{\eta^2}\right) = 0\,.
\end{equation}
The solutions of equation (\ref{GWinTGnodampingQuasideSitter}) are given in terms of Hankel functions
\begin{equation}\label{solutionsTG}
    g(\eta,k) \subset \left\lbrace C(k)\sqrt{|\eta|}H_{\alpha}^{(1)}(k|\eta|), \quad C^{*}(k)\sqrt{|\eta|}H_{\alpha}^{(2)}(k|\eta|) \right\rbrace, \quad \alpha = \frac{3}{2} + \epsilon(1+ \gamma)\,,
\end{equation}
such that the initial condition requires the choosing of the first solution, wherein the infinite past implies the integration constant to be
\begin{equation}
    C(k) = \frac{\sqrt{\pi}}{2}e^{i\alpha\frac{\pi}{2} + i\frac{\pi}{4}}\,,
\end{equation}
where $\alpha$ is given in Eq.~\ref{solutionsTG}.
The form of the solutions is completely analogous to those of GR \citep{Piattella2018} but with a different $\alpha$ for the Hankel functions, due to the presence of $\gamma$. If $\gamma=0$ we recover the same solution as GR, which is expected. In the case of a quasi de Sitter, we have
\begin{equation}
    f(\eta) = \frac{1}{a(\eta)}\text{Exp}\left(-\frac{1}{2}\int \mathcal{H}\nu d\eta \right) = a(\eta)^{-(1+ \epsilon \gamma)}\,,
\end{equation}
from where the original $h$ field is
\begin{equation}
    \hat h = a^{-(1+ \epsilon \gamma)}\frac{\sqrt{\pi}}{2}e^{i \frac{\alpha \pi }{2}  +  i\frac{\pi}{4}}\sqrt{|\eta|}H_{\alpha}^{(1)}(k|\eta|)\,.
\end{equation}
The power spectrum is then given by
\begin{equation}\label{PowerSpectrumTG}
    P_h(\eta,k) = 32\pi G a^{-2(1 + \epsilon \gamma)} \frac{\pi}{4}|\eta||H_{\alpha}^{(1)}(k|\eta|)|^2\,,
\end{equation}
and the dimensionless power spectrum is given by
\begin{equation}
    \Delta^2_h(\eta, k) = \frac{k^3 |\eta|}{\pi M_{\text{pl}}^2} a^{-2(1 + \epsilon \gamma)}|H_{\alpha}^{(1)}(k|\eta|)|^2\,.
\end{equation}
In the superhorizon limit, the Hankel function has the following asymptotic behavior
\begin{equation}
    H_{\alpha}^{(1)}(x) \underset{x \to 0}{\sim}-i \frac{\Gamma(\alpha)}{\pi}\left(\frac{x}{2}\right)^{-\alpha}\,,
\end{equation}
from where the dimensionless power spectrum, using the explicit form of the scale factor, becomes
\begin{equation}
    \Delta^{2}_h(\eta,k) =  \frac{k^3 |\eta|\Gamma(\alpha)^2}{\pi^3 M_{\text{pl}}^2} H_{\Lambda}^{2(1 + \gamma\epsilon)}|\eta|^{2 + 2\epsilon(1+ \gamma)} \left(\frac{k|\eta|}{2}\right)^{-3 - 2\epsilon(1+\gamma)} = \frac{\Gamma(\alpha)^2 2^{3 + 2\epsilon(1+\gamma)}}{\pi^3M_{\text{pl}}^2}H_{\Lambda}^{2(1 + \gamma\epsilon)}k^{-2\epsilon(1 + \gamma)}\,,
\end{equation}
such that taking the $k$-dependence at first order in $\epsilon$ and zero-order everywhere else, we obtain 
\begin{equation}
    \Delta^{2}_h(\eta,k) = \frac{2 H_{\Lambda}^2}{\pi^2 M_{\text{pl}}^2 }k^{-2\epsilon(1 + \gamma)}\,.
\end{equation}
The power law dependence on the scale is called \textit{tensor spectral index}, where in the case of TG reads as
\begin{equation}
    n_T = -2\epsilon(1 + \gamma)\,,
\end{equation}
which has an extra contribution $\gamma$ due to the presence of extensions of TEGR. This is a key result since if the GW background is measured and a large tensor spectral index is measured, it would suggest the need for extensions of gravity, since in GR, $n_T = -2\epsilon$, and then must be small consequently.

\end{itemize}


\subsection{Power spectrum including tensor anisotropies}
\label{subsec:PS3}

For this section, we will follow a similar approach as in Ref.~\citep{Klose2022} but for a general source of tensor anisotropic stress. The idea is to decompose the tensor field into two modes, a short wavelength mode, and a slowly variant long wavelength mode
\begin{equation}\label{FieldSplit}
    h(\eta,\boldsymbol{x}) = h_{<}(\eta,\boldsymbol{x}) + h_{>}(\eta,\boldsymbol{x})\,.
\end{equation}
The short wavelength part contains all the information of quantum vacuum fluctuations and its effects are manifested through the parameters of the long wavelength perturbation, in this case, through the power spectrum. The short wavelength mode represents high momentum modes with wavelength much smaller than the horizon, We then write the short wavelength mode as a quantized field but with a filter or windows function
\begin{equation}\label{ShortWavelengthMode}
    h_{<}(\eta,\boldsymbol{x})= \int \frac{d^3 \boldsymbol k}{\sqrt{(2\pi)^3}}W(\eta,k)\left[h(\eta,k)e^{i\mathbf k\cdot\mathbf x}a(\mathbf k,\lambda)+ h^*(\eta,k)e^{-i\mathbf k\cdot\mathbf x}a^{\dagger}(\mathbf k,\lambda)\right]\,,
\end{equation}
where the filter function is given by
\begin{equation}
    W(\eta,k) = \theta(k - \varepsilon \mathcal{H})\,,
\end{equation}
and a suitable parameter $\varepsilon$ has been introduced such that it allows us to guarantee that $h_{<}$ is a short wavelength contribution. Taking the GW equations of TG in a vacuum and the physical space for a quasi-de Sitter background, we have
\begin{equation}
    h'' - \frac{2}{\eta}\left[1 + \epsilon(1+ \gamma) \right]h' - \nabla^2 h = 0\,,
\end{equation}
such that, if we perform the split (\ref{FieldSplit}), we arrive at
\begin{equation}\label{LongEquation}
    h''_{>} - \frac{2}{\eta}\left[1 + \epsilon(1+ \gamma) \right]h'_{>} - \nabla^2 h_{>} = \varrho_{Q}, \quad \varrho_{Q} = - \left(h''_{<} - \frac{2}{\eta}\left[1 + \epsilon(1+ \gamma) \right]h'_{<} - \nabla^2 h_{<} \right)\,,
\end{equation}
with $\varrho_Q$ is called the \textit{quantum noise}. In the case of TG in a quasi-de Sitter background, the quantum noise has the following form 
\begin{equation}
    \varrho_Q(\eta,\boldsymbol{x}) = - \int \frac{d^3 \boldsymbol k}{\sqrt{(2\pi)^3}}\left[a(\mathbf k,\lambda)f^{\text{TG}}_{k}(\eta)e^{i\mathbf k\cdot\mathbf x} + a^{\dagger}(\mathbf k,\lambda){f^{\text{TG}}_{k}}^*(\eta)e^{-i\mathbf k\cdot\mathbf x}  \right]\,,
\end{equation}
with 
\begin{equation}
    f^{\text{TG}}_{k}(\eta) = \left(W'' - \frac{2}{\eta}\left[1 + \epsilon(1 + \gamma) \right]W'\right)h + 2W'h'\,.
\end{equation}
In the limit $\epsilon \to 0$ we recover the same quantum noise found in ref. \citep{Klose2022b}. The correlator for the quantum noise is 
\begin{equation}
    \bra{0}\varrho_Q(\eta_1,\boldsymbol{x}_1)\varrho_Q(\eta_2,\boldsymbol{x}_2)\ket{0} = \int \frac{d^3 \boldsymbol k}{(2\pi)^3}e^{i \boldsymbol k \cdot(\boldsymbol{x}_1 - \boldsymbol{x}_2)}f_k(\eta_1)f^*_{k}(\eta_2)\,,
\end{equation}
or in Fourier space is
\begin{equation} \label{ExpectationFourierQuantumNoise}
    \bra{0}\varrho_Q(\eta_1,\boldsymbol{k})\varrho_Q(\eta_2,\boldsymbol{q})\ket{0}  = \delta_D(\boldsymbol k - \boldsymbol q)f_k(\eta_1)f^*_{k}(\eta_2)\,.
\end{equation} 
However, we want to include explicitly the contributions of tensor anisotropic stress, for that, we consider the propagation equation for GW in TG
\begin{equation}
    \hat h''_{kj} + [2 + \nu]\mathcal{H}\hat h'_{kj}  + k^2 \hat h_{kj} = \frac{16\pi G a^{2}}{f_T}\pi_{ij}^T\,,
\end{equation}
and then multiply both sides by $\epsilon_{ij}^{\lambda *}$, sum over the spatial indices and use the orthonormality (\ref{Orthonormality}) of the polarization tensors to arrive at
\begin{equation}
    \left(\partial^2_{\eta} + [2 + \nu]\mathcal{H}\partial_{\eta} + k^2 \right)\hat{h}^{\lambda}(\eta,\boldsymbol k) = \frac{16\pi G a^{2}}{f_T}\sum_{ij}\epsilon_{ij}^{\lambda *}\pi_{ij}^T \equiv -\frac{1}{f_T}\varrho_{T}^{\lambda}(\eta,\boldsymbol k)\,,
\end{equation}
with $\varrho_{T}^{\lambda}(\eta,\boldsymbol k)$ is the tensor anisotropic noise. If $f_T = -1$ we obtain the same equation as in ref. \citep{Klose2022b}. Now, we perform the split into short and long wavelength modes (\ref{FieldSplit}), leaving us with the following equation
\begin{equation}
    \left(\partial^2_{\eta} + [2 + \nu]\mathcal{H}\partial_{\eta} + k^2 \right)\hat{h}^{\lambda}_{>}(\eta,\boldsymbol k) = \hat \varrho_{Q}^{\lambda}(\eta,\boldsymbol k) -\frac{1}{f_T}\varrho_{T}^{\lambda}(\eta,\boldsymbol k)\,,
\end{equation}
where the hat in $\hat \varrho_{Q}^{\lambda}(\eta,\boldsymbol k)$ is to indicate that it is computed using the canonical normalization of the tensor perturbation. Now, we perform the same change of variable as before
\begin{equation}
    \hat h_{>}^{\lambda}(\eta,\boldsymbol k) = f(\eta)g^{\lambda}_{>}(\eta,\boldsymbol k) \quad \text{with} \quad f(\eta) = a(\eta)^{-(1 + \gamma \epsilon)}\,,
\end{equation}
such the equation for GW in a quasi de Sitter becomes
\begin{equation}\label{EqTensorContribution}
    \left(\partial^2_{\eta} + k^2 - \frac{2 + 3\epsilon(1+\gamma)}{\eta^2} \right)g^{\lambda}_{>}(\eta,\boldsymbol k) =  a(\eta)^{1 + \gamma \epsilon}\left(\hat \varrho_{Q}^{\lambda}(\eta,\boldsymbol k) -\frac{1}{f_T}\varrho_{T}^{\lambda}(\eta,\boldsymbol k) \right)\,.
\end{equation}
We will solve this equation using the method of Green's function with retarded boundary conditions, such that 
 \begin{equation}
     \left(\partial^2_{\eta} + k^2 - \frac{2 + 3\epsilon(1+\gamma)}{\eta^2} \right)G_{R}(\eta,\eta_i,k) = \delta_D(\eta - \eta_i)\,,
\end{equation}
and $G_R(\eta,\eta_i,k)=0$ if $\eta < \eta_i$. The Green's function is computed from the linear independent solution from the vacuum case in (\ref{solutionsTG}) from where it reads as
\begin{equation}\label{Green's TG}
    G_{\text{TG}}^R(\eta, \eta_i,k) = \theta(\eta - \eta_i)\frac{\pi}{2}\sqrt{|\eta|}\sqrt{|\eta_i|}\text{ Im}\left[H_{\alpha}^{(1)}(k |\eta_i|)H_{\alpha}^{(2)}(k |\eta|) \right], \quad \alpha = \frac{3}{2} + \epsilon(1 + \gamma)\,.
\end{equation}
The solution of the equation is then 
\begin{equation}
    g^{\lambda}_{>}(\eta,\boldsymbol{k})  = \int_{-\infty}^{\eta}d\eta_i a(\eta_i)^{1 + \gamma \epsilon}G_{\text{TG}}^R(\eta, \eta_i,k) \hat \varrho_Q^{\lambda}(\eta_i,\boldsymbol{k}) - \int_{-\infty}^{\eta}d\eta_i \frac{a(\eta_i)^{1 + \gamma \epsilon}}{f_T(\eta_i)}G_{\text{TG}}^R(\eta, \eta_i,k) \varrho_T^{\lambda}(\eta_i,\boldsymbol{k})\,.
\end{equation}
If we choose $\eta_i$ to be the time when the mode enters the horizon, we can interpret the retarded Green's function as the Green's function with boundary conditions such that when the mode has not re-entered the horizon yet, the long wavelength mode is zero and we end up with only the short-wavelength mode. Returning to the $h$ field, we obtain the general solution for GW in TG in a quasi-de Sitter background that includes the effects of tensor anisotropies
\begin{align}\label{GeneralSolutionTG}
    \hat h^{\lambda}_{>}(\eta,\boldsymbol{k}) =& \int_{-\infty}^{\eta}d\eta_i \left[\frac{a(\eta_i)}{a(\eta)}\right]^{1 + \gamma \epsilon}G_{\text{TG}}^R(\eta, \eta_i,k) \hat \varrho_Q^{\lambda}(\eta_i,\boldsymbol{k})  - \int_{-\infty}^{\eta}d\eta_i \frac{1}{f_T(\eta_i)}\left[\frac{a(\eta_i)}{a(\eta)}\right]^{1 + \gamma \epsilon}G_{\text{TG}}^R(\eta, \eta_i,k) \varrho_T^{\lambda}(\eta_i,\boldsymbol{k})\,. 
\end{align}
This result is of high importance since it opens the door to studying widely different possible scenarios of generations of GW in the context of TG. if $f_T = -1$, then $\gamma = 0$, and then we have also the general solution in GR for a quasi-de Sitter background. In the limit $\epsilon \to 0$, the Green's function satisfy 
\begin{equation}
    \frac{a(\eta_i)}{a(\eta)}G^{R}(\eta,\eta_i,k) = \frac{\theta(\eta - \eta_i)}{\eta_i^2 k^3}\text{Im}\left[e^{ik(\eta - \eta_i)}(1 - ik\eta)(1 + ik\eta_i) \right]\,,
\end{equation}
which is the same Green's function found in \citep{Klose2022b}.
We are now in a position to see the form of the power spectrum using this result. Taking the equal time correlator of the general solution, we have

\begin{align} 
    &\left\langle \hat h^{\lambda}_{>}(\eta,\boldsymbol{k}) \hat h^{\lambda '}_{>}(\eta,\boldsymbol{q}) \right\rangle = \\ & \int_{-\infty}^{\eta} d\eta_{1}\left[\frac{a(\eta_1)}{a(\eta)}\right]^{1 + \gamma \epsilon}G_{\text{TG}}^R(\eta, \eta_1,k)\int_{-\infty}^{\eta} d\eta_{2}\left[\frac{a(\eta_2)}{a(\eta)}\right]^{1 + \gamma \epsilon}G_{\text{TG}}^R(\eta, \eta_2,k)\bra{0}\hat \varrho_Q^{\lambda}(\eta_1,\boldsymbol{k})\hat \varrho_Q^{\lambda '}(\eta_2,\boldsymbol{q})\ket{0} \notag \\ & \notag + \int_{-\infty}^{\eta} d\eta_{1}\frac{1}{f_T(\eta_1)}\left[\frac{a(\eta_1)}{a(\eta)}\right]^{1 + \gamma \epsilon}G_{\text{TG}}^R(\eta, \eta_1,k)\int_{-\infty}^{\eta} d\eta_{2}\frac{1}{f_T(\eta_2)}\left[\frac{a(\eta_2)}{a(\eta)}\right]^{1 + \gamma \epsilon}G_{\text{TG}}^R(\eta, \eta_2,k) \\ \notag & \bra{0} \varrho_T^{\lambda}(\eta_1,\boldsymbol{k}) \varrho_T^{\lambda '}(\eta_2,\boldsymbol{q})\ket{0}\,,
\end{align}
where recalling the correlator of the quantum noise
\begin{equation}
    \bra{0}\hat \varrho_Q^{\lambda}(\eta_1,\boldsymbol{k})\hat \varrho_Q^{\lambda '}(\eta_2,\boldsymbol{q})\ket{0} = 32\pi G \delta_{D}(\boldsymbol k - \boldsymbol q)\delta^{\lambda \lambda'}f_{k}^{\text{TG}}(\eta_1){f_{k}^{\text{TG}}}^*(\eta_2)\,,
\end{equation}
and taking into account the two polarization states implies that
\begin{eqnarray} \label{CorrelatorGeneral}
    && \left\langle \hat h_{>}(\eta,\boldsymbol{k}) \hat h_{>}(\eta,\boldsymbol{q}) \right\rangle =  64\pi G \delta_{D}(\boldsymbol k - \boldsymbol q) \left\lvert\int_{-\infty}^{\eta}d\eta_{i}\left[\frac{a(\eta_i)}{a(\eta)}\right]^{1 + \gamma \epsilon}G_{\text{TG}}^R(\eta, \eta_i,k)f_{k}^{\text{TG}}(\eta_i) \right\rvert^2 \\ \notag   
    && + \int_{-\infty}^{\eta} d\eta_{1}\frac{1}{f_T(\eta_1)}\left[\frac{a(\eta_1)}{a(\eta)}\right]^{1 + \gamma \epsilon}G_{\text{TG}}^R(\eta, \eta_1,k)\int_{-\infty}^{\eta} d\eta_{2}\frac{1}{f_T(\eta_2)}\left[\frac{a(\eta_2)}{a(\eta)}\right]^{1 + \gamma \epsilon}G_{\text{TG}}^R(\eta, \eta_2,k) \sum_{\lambda}\bra{0} \varrho_T^{\lambda}(\eta_1,\boldsymbol{k}) \varrho_T^{\lambda }(\eta_2,\boldsymbol{q})\ket{0}\,.
\end{eqnarray}
By construction, before the mode enter the horizon $-\eta  < - \eta_*$, with $\eta_*$ the time of horizon crossing, the term 
\begin{equation}
    P_{\Lambda}(\eta,k) =  64\pi G \left\lvert\int_{-\infty}^{\eta}d\eta_{i}\left[\frac{a(\eta_i)}{a(\eta)}\right]^{1 + \gamma \epsilon}G_{\text{TG}}^R(\eta, \eta_i,k)f_{k}^{\text{TG}}(\eta_i) \right\rvert^2 =\frac{2\pi |\eta|}{ M_{\text{pl}}^2} a^{-2(1 + \epsilon \gamma)}|H_{\alpha}^{(1)}(k|\eta|)|^2\,,
\end{equation}
which is the contribution from the vacuum computed previously in (\ref{PowerSpectrumTG}) considering both polarization states. Analogously, we define the contribution to the power spectrum from tensor anisotropic stress as
\begin{align}
    \delta_{D}(\boldsymbol k - \boldsymbol q)P_{T}(\eta,k)=&\int_{-\infty}^{\eta} d\eta_{1}\frac{1}{f_T(\eta_1)}\left[\frac{a(\eta_1)}{a(\eta)}\right]^{1 + \gamma \epsilon}G_{\text{TG}}^R(\eta, \eta_1,k)\int_{-\infty}^{\eta} d\eta_{2}\frac{1}{f_T(\eta_2)}\left[\frac{a(\eta_2)}{a(\eta)}\right]^{1 + \gamma \epsilon}G_{\text{TG}}^R(\eta, \eta_2,k)\\ \notag & \sum_{\lambda}\bra{0} \varrho_T^{\lambda}(\eta_1,\boldsymbol{k}) \varrho_T^{\lambda}(\eta_2,\boldsymbol{q})\ket{0}\,.
\end{align} 
There exist possible sources for anisotropic stress studied in the context of GR, for instance, local thermal fluctuations of the primordial plasma \citep{Klose2022b}, first-order cosmological phase transitions \citep{Hogan1986,Durrer2010,Grojean2007}, and magnetic fields \citep{Caprini2006}, to name a few. Whether or not those physical mechanisms sourcing GW are present in TG is a matter of future research, however, at this point, it is possible to discuss the implication of the extended models on those peaks on the power spectrum, as we shall see later on.


\section{Energy density of gravitational waves in teleparallel gravity}
\label{sec:GWenergy}

The energy density spectrum is defined as the gravitational-wave energy density per
logarithmic wave number interval is given by \citep{Boyle2008}
\begin{equation}\label{EnergySpectrumGW}
    \Omega_{\text{GW}}(\eta,k) = \frac{1}{\rho_{\text{crit}}(\eta)} \frac{\bra{0}\rho_{\text{GW}}(\eta) \ket{0} }{d \ln k}\,,
\end{equation}
where the critical density is given by
\begin{equation}
    \rho_{\text{crit}}(\eta) = \frac{3H^2(\eta)}{8\pi G}\,.
\end{equation}
To compute $\rho_{\text{GW}}$ we will compute the energy-momentum tensor associated with the action for GW in TG, which is given by \citep{Bahamonde:2021gfp}
\begin{equation}
    S = \int d^3\boldsymbol{x}d\eta (-a^2 f_T)\left[ h'_{ij}h'^{ij} - \partial_{k}h_{ij}\partial^{k}h^{ij} \right]\,,
\end{equation}
or in terms of the background tetrad
\begin{equation}
    S = \int d^3\boldsymbol{x}d\eta \bar e(-f_T)\left[\bar E^{\mu}_{ \ A}\bar E^{\nu}_{ \ B}\eta^{AB} \partial_{\mu}h_{ij}\partial_{\nu}h_{ij}\right]\,,
\end{equation}
where it is possible to identify the Lagrangian density to be
\begin{equation}
    \mathcal{L}= (-f_T)\left[\bar E^{\mu}_{ \ A}\bar E^{\nu}_{ \ B}\eta^{AB} \partial_{\mu}h_{ij}\partial_{\nu}h_{ij}\right]\,.
\end{equation}
With this Lagrangian density, it is possible to compute the associated energy-momentum tensor given by eq. (\ref{Energy-momentum tensor TG}), taking into account the signature $(+,-,-,-)$, whose result is
\begin{equation}
    \mathcal{T}^{\lambda}_{\ \nu} = 2(-f_T)\bar E^{\lambda}_{\ A}\bar E^{\mu}_{\ B}\eta^{AB}\partial_{\mu}h_{ij}\partial_{\nu}h_{ij} + \left[\bar E^{\mu}_{ \ C}\bar E^{\alpha}_{ \ B}\eta^{CB} \partial_{\mu}h_{ij}\partial_{\alpha}h_{ij}\right]\bar e^{A}_{\ \nu}\frac{1}{\bar e}W^{\lambda}_{\ A}[f_T,\bar e]\,,
\end{equation}
where we have defined
\begin{align}
    W^{\lambda}_{\ A}[f_T,\bar e]\equiv  \frac{\delta(\bar e f_T(T,B))}{\delta \bar e^{A}_{\ \lambda}} =& -\bar ef_{TB}\bar B \bar E^{\lambda}_{A} - 2\bar e \bar E^{\lambda}_{A}\accentset{\circ}{\square}f_{TB} +  2\bar e \bar E^{\nu}_{A}\accentset{\circ}{\bar \nabla}^{\lambda}\accentset{\circ}{\bar \nabla}_{\nu}f_{TB} - 4\bar e(\partial_{\mu}f_{TB})\bar S_A^{\text{   }\mu \lambda} \\ \notag &- 4\partial_{\mu}(f_{TT})\bar e\bar S_{A}^{\text{   }\mu \lambda} - 4f_{TT} \partial_{\mu}\left(\bar e\bar S_{A}^{\mu \lambda} \right) + 4\bar ef_{TT} \bar T^{\alpha}_{\text{   }\mu A}\bar S_{\alpha}^{\text{   }\lambda \mu} + \bar ef_T\bar E^{\lambda}_A\,,
\end{align}
to keep the same notation as the case of the field equations and we have neglected total derivatives. Thus, we identify
\begin{equation}
    \rho_{\text{GW}}^{\text{TG}}=\Theta^{0}_{\ 0}  = \frac{2}{a^2(\eta)}\left[ -(h'_{ij})^2(f_T + \kappa \bar \rho_{f_T}) + \kappa \bar \rho_{f_T}(\nabla h_{ij})^2 \right]\,, \label{TGrho}
\end{equation}
with 
\begin{equation}
    a^2(\eta)\kappa \bar \rho_{f_T} \equiv 3 \mathcal{H}f'_{TB} - 6\mathcal{H}^2(f_{TB} + f_{TT}) - 3f_{TB}\mathcal{H}' - \frac{a^2(\eta)}{2}f_T\,.
\end{equation}
Notice that $f_T=-1$ implies $\kappa \bar \rho_{f_T}=\frac{1}{2}$ and we recover the GR limit \citep{Boyle2008}. Using the condition at horizon crossing \citep{Boyle2008}
\begin{equation}
    |h'(\eta,k)|^2 = k^2 |h(\eta,k)|^{2}\,, 
\end{equation}
and going to Fourier space, we obtain
\begin{equation}
    \Omega(\eta,k) = \frac{1}{12}\frac{k^2 \Delta_h^2(\eta,k)}{a^2(\eta)H^2(\eta)}\left(-f_T\right)\,,
\end{equation}
where the GR result is trivially recovered by $f_T=-1$. 


\section{Discussions}
\label{sec:discussions}

There are several implications of the main results of this work. On the one hand, when working in a vacuum for a perfect de Sitter background, both GW in GR and TG satisfy the same propagation equation (\ref{GREquivalentEquation}), and therefore, have the same analytical solution. This implies that the associated scale-invariant power spectrum is the same for both theories. Thus, if the GW background is measured with zero or nearly zero tensor spectral index (scale-invariant power spectrum), both theories will survive, and there are no direct implications of the plausibility of TG as the underlying gravitational theory. 

On the other hand, in the case when the background is promoted to a quasi de Sitter background, driven by the first slow roll parameter $\epsilon$, the situation is quite different. As we have seen, the $\gamma$ parameter encodes all the information about the presence of extensions of gravity. In the case of a quasi de Sitter background, the tensor spectral index is $n_T = -2\epsilon(1+\gamma)$. Hence, if a future GW detector is capable of directly measuring the tensor spectral index and observes a value way different from zero, this would indicate a need of extensions of gravity since TG allows for high values of the tensor scalar index. This is an important result and relies on the efficacy of future GW detectors for measures of the GW background, since the only way we have to infer the tensor spectral index is through the tensor-to-scalar ratio, and this ratio is highly dependent on the gravitational theory.

Therefore, a direct measurement of zero tensor spectral index does not provide enough evidence to discard this expression of TG, but a measurement of a tensor spectral index way higher than zero is a direct implication of the need for an extended gravitational model, where we have shown that such result is perfectly explained by TG, providing a significant prediction and not only a validation. 

Furthermore, if we take the limit $\epsilon \to 0$, the $f(T,B)$ functional will not depend on time $f'_T = 0$, and Green's function is given in Eq.~(\ref{Green's TG}) will recover that of GR 
\begin{equation}
     \frac{a(\eta_i)}{a(\eta)}G^{R}(\eta,\eta_i,k) = \frac{\theta(\eta - \eta_i)}{\eta_i^2 k^3}\text{Im}\left[e^{ik(\eta - \eta_i)}(1 - ik\eta)(1 + ik\eta_i) \right] \equiv G_{R}(\eta,\eta_i,k)\,,
\end{equation}
using the same notation as in \citep{Klose2022b}. This implies that the tensor contribution to the power spectrum can be written as
\begin{align}
     \delta_{D}(\boldsymbol k - \boldsymbol q)P_{T}(\eta,k)=&\frac{1}{f_T^2}\int_{-\infty}^{\eta} d\eta_{i}G_{R}^2(\eta,\eta_i,k)\sum_{\lambda}\bra{0} \varrho_T^{\lambda}(\eta_i,\boldsymbol{k}) \varrho_T^{\lambda}(\eta_i,\boldsymbol{q})\ket{0} \equiv \frac{1}{f_T^2}  \delta_{D}(\boldsymbol k - \boldsymbol q)P_T^{GR}(\eta,k)\,,
\end{align} 
with
\begin{equation}
    \delta_{D}(\boldsymbol k - \boldsymbol q)P_T^{GR}(\eta,k) = \int_{-\infty}^{\eta} d\eta_{i}G_{R}^2(\eta,\eta_i,k)\sum_{\lambda}\bra{0} \varrho_T^{\lambda}(\eta_i,\boldsymbol{k}) \varrho_T^{\lambda}(\eta_i,\boldsymbol{q})\ket{0}\,,
\end{equation}
the contribution from tensor anisotropies on the power spectrum. Therefore, we will have either an amplification on the peaks of the power spectrum, if $-1 <f_T <0$, or a reduction, if $f_T <-1$, coming from tensor anisotropies. For particular models studied in the literature, like the ones we presented previously in Sec.~\ref{subsec:PS2}, the functional satisfies $f_T = -1 + F_T$ with $F_T <0$, hence $f_T <-1$. This would imply that the peaks from tensor anisotropies are squeezed in those particular models, giving us another important observational prediction. If the peaks on the power spectrum, from tensor anisotropies, of primordial GW are decreased compared to what should be expected from GR, it is a hint of the need for TG as a possible gravitational theory, where the amount of squeezing will depend highly on the particular choice of the $(T,B)$ functional. Finally, on the same $\epsilon \to 0$ limit and  $f_T = -1 + F_T$ with $F_T <0$, it is possible to observe that 
\begin{equation}
    \Omega(\eta,k) = \frac{1}{12}\frac{k^2 \Delta_h^2(\eta,k)}{a^2(\eta)H^2(\eta)} - F_T\frac{1}{12}\frac{k^2 \Delta_h^2(\eta,k)}{a^2(\eta)H^2(\eta)}\,,
\end{equation}
and identifying the contribution from GR and the contribution from extensions of gravity 
\begin{equation}
    \Omega_{\text{GR}} \equiv  \frac{1}{12}\frac{k^2 \Delta_h^2(\eta,k)}{a^2(\eta)H^2(\eta)}\,, \quad \Omega_{F_T} = - F_T\frac{1}{12}\frac{k^2 \Delta_h^2(\eta,k)}{a^2(\eta)H^2(\eta)}\,,
\end{equation}
we have 
\begin{equation}
    \Omega(\eta,k)  = \Omega_{\text{GR}} + \Omega_{F_T}  > \Omega_{GR}\,,
\end{equation}
this implies that for the particular cosmological viable models of $f(T, B)$ presented in subsection \ref{subsec:PS2}, the amount of energy density in TG is greater than that of GR, which is also an important prediction for TG since if the energy spectrum observed is greater than the expected from GR, the need for an extension of gravity will be manifest with $f(T,B)$ or $f(T)$ being a candidate to explain that extra energy density. 

The bigger energy density of GW in TG, along with the results of reduction in the peaks of the power spectrum, implies that the vacuum fluctuations have a bigger contribution to the energy density of GW in TG compared with its contribution in GR, and the contributions from tensor anisotropies have less contribution.


\section{Conclusions}
\label{sec:conclusions}

TG is a gravitational theory built as a gauge theory of translations and invariance under local Lorentz transformations. In its equivalent extension limit to TEGR, TG is completely equivalent to GR, hence showing the same advantages and disadvantages in explaining the cosmological phenomena. The need for extended models is an important subject of research in TG to try to solve the GR problems which are manifest in the TEGR limit. These models have been explored widely in the context of the late universe to address the cosmological constant issue, cosmography, and other problems, and some contributions in the context of the early universe have been performed, including studies of primordial black holes \citep{ElBourakadi2022} and inflation \citep{Rezazadeh2016,Sahlu2020,Chakrabortty2021}. As a further contribution, in this paper, we study the production of primordial GW in the context of TG. We showed that in a perfect de Sitter background the propagation equation in TG is the same as in the GR case, then, the same signal as GR of the power spectrum is recovered in TG since it is given from the squared modulus of the solution of the same equation for GW. In the quasi de Sitter limit we obtained that TG allows for a high value of the tensor spectral index, providing a crucial test of TG if a direct measurement of the power spectrum of GW from vacuum fluctuations is done and a tensor spectral index obtained is considerably different from zero, this would imply the need of TG. It is interesting to notice that our results could be tested via future data from pulsar timing arrays, e.g. PTAs. In particular, such experiments have constrained some effects related to stochastic gravitational wave backgrounds as sources of cosmic inflation \citep{Benetti2022}. A successful program through these experiments could determine if we will observe a blue tilt on the tensor power spectrum. Finally, we obtained that although the vacuum contribution in a perfect de Sitter background is the same for both GR and TG. In both cases, the peaks of the contributions from tensor anisotropies are different, particularly, we explored these differences for some particular cosmological viable models, showing that the peaks are squeezed in TG compared to GR. 

Other works on the topic of gravitational waves induced from primordial black holes include Refs.~\citep{Papanikolaou:2022hkg,2303.16695} which tackle different aspects of the topic. In the present work, we focus on the generation of primordial gravitational waves, in the context of linear perturbation theory, coming from vacuum fluctuations and tensor anisotropic stress which is a first order perturbation quantity of the energy momentum tensor and do not address the generation of gravitational waves induced from second order scalar perturbation. Secondly, and the most important one, primordial black holes are formed in the radiation dominated era, and in this work, we go far earlier and address the generation during the inflaton dominated era where the universe stays in a quantum state.

Finally, we showed that we expect a greater amount of energy density in TG compared to GR. This result implies that along with a reduction in the peaks of the power spectrum, the vacuum fluctuations have a greater contribution to the energy density in TG and the tensor anisotropies have a less contribution when compared to GR. These results are the basis for the study of primordial GW in TG, but to perform a complete numerical analysis that involves cosmological constraints, we need to compute the transfer functions in TG and study the characteristic signals that should be expected within the redshift range of the GW experiment, and then use the data to test both theories.

Within the same latter scheme, we included tensor anisotropies sourcing GW, and the direct observational implications on the peaks of the power spectrum were computed. This is an important step in studying TG in the context of the early universe. In the case of only vacuum contributions, the result for the power spectrum needs to be taken in the superhorizon limit and evaluated at the time of horizon crossing when the $k$ mode re-enters the horizon.  In the case of including tensor anisotropies, it should be evaluated at a time $\tilde \eta$ shortly after the end of inflation, and it will depend on the phenomenology of the tensor source. Nevertheless, these expressions need to be studied at cosmic times (redshifts) achievable by future experiments, and that will require the study of transfer functions in TG which will allow us to evaluate these results at the required scales. However such a task lies beyond the scope of this paper, but in the light of future GW detectors, like \href{https://lisa.nasa.gov/}{LISA} or the \href{https://www.et-gw.eu}{Einstein Telescope}, this should bring the attention of the TG community to explore these topics with the general solutions provided here.


\bigskip

\section*{Acknowledgmenents}
GRF is supported by CONACyT National Grant. CE-R acknowledges the Royal Astronomical Society as FRAS 10147 and is supported by PAPIIT UNAM Project TA100122. This article is based upon work from COST Action CA21136 Addressing observational tensions in cosmology with systematics and fundamental physics (CosmoVerse) supported by COST (European Cooperation in Science and Technology).


\appendix
\section{Tensor perturbations}
The idea of tensor perturbation consists in splitting the tetrad field into two arts, a background tetrad, referring to the background manifold exhibiting cosmological symmetry $\bar{e}^{A}_{\ \mu}$, and a small deviation from this background $\delta e^{A}_{\ \mu}$ such that
\begin{equation}
    e^{A}_{\ \mu} = \bar{e}^{A}_{\ \mu} + \delta e^{A}_{\ \mu}. \label{eq:LinearP}
\end{equation}
We can identify the geometrical part of field equations \eqref{f(T,B) field equations} as 
\begin{align}
W^{\lambda}_{\nu} = \delta^{\lambda}_{\nu} A - B^{\lambda}_{\ \nu} + \frac{1}{2}\delta^{\lambda}_{\nu} C + 2D^{\lambda}_{\ \nu} + 2 F^{\lambda}_{\ \nu} -2H^{\lambda}_{\ \nu}  - \frac{1}{2}f\delta^{\lambda}_{\nu}, \label{eq:FieldEqR}
\end{align}
with 
\begin{eqnarray}
&& A= \accentset{\circ}{\square}f_B, 
\quad
B^{\lambda}_{\ \nu} = \accentset{\circ}{\nabla}^{\lambda}\accentset{\circ}{\nabla}_{\nu}f_B, 
\quad 
C = f_B B, 
\quad D^{\lambda}_{\ \nu}=\left[\partial_{\mu}f_B + \partial_{\mu}f_T \right]S_{\nu}^{\text{   }\mu \lambda}, 
\\
&& F^{\lambda}_{\ \nu} = \frac{1}{e}e^{A}_{\ \nu}f_T \partial_{\mu}(e S_{a}^{\ \mu \lambda}), 
\quad
H^{\lambda}_{\ \nu} = f_T T^{\alpha}_{\ \mu \nu}S_{\alpha}^{\ \lambda \mu}. \label{174}
\end{eqnarray}
The linear perturbation of the geometrical part of field equations \eqref{eq:FieldEqR} due to the linear perturbation \eqref{eq:LinearP} is

\begin{align}\label{perturbedGTG}
\delta W^{\lambda}_{\nu} = \delta^{\lambda}_{\nu} \delta A - \delta B^{\lambda}_{\ \nu} + \frac{1}{2}\delta^{\lambda}_{\nu} \delta C + 2\delta D^{\lambda}_{\ \nu} + 2 \delta F^{\lambda}_{\ \nu} -2\delta H^{\lambda}_{\ \nu}  - \frac{1}{2}\delta f\delta^{\lambda}_{\nu},
\end{align}
with 
\begin{align}
\delta A &=  \delta g^{\mu \beta} \left(\partial_{\mu}\partial_{\beta}\bar f_{\bar B} - \accentset{\circ}{\bar \Gamma}^{\rho}_{\beta \mu}\partial_{\rho} \bar f_{\bar B}\right) + \bar g^{\mu \beta}\left(\partial_{\mu}\partial_{\beta} \delta f_B  - \delta \accentset{\circ}{\Gamma}^{\rho}_{\beta \mu}\partial_{\rho} \bar f_{\bar B} -  \accentset{\circ}{\bar \Gamma}^{\rho}_{\beta \mu}\partial_{\rho}\delta f_B \right), \\
\delta B^{\lambda}_{\ \nu} &=  \delta g^{\lambda \beta} \accentset{\circ}{\bar \nabla}_{\beta}\accentset{\circ}{\bar \nabla}_{\nu}\bar f_{\bar B}  +  \bar  g^{\lambda \beta}\left(  \accentset{\circ}{\bar \nabla}_{\beta}\accentset{\circ}{\bar \nabla}_{\nu}\delta f_B - \delta \accentset{\circ}{\Gamma}^{\rho}_{\nu \beta}\partial{\rho}\bar f_{\bar B}\right), \\
\delta C &=\bar  f_{\bar B} \delta B +  \delta  f_B \bar B, \\
\delta D^{\lambda}_{\ \nu} &= \partial_{\mu}\left(\delta f_B  +  \delta f_T  \right)\bar S_{\nu}^{\ \mu \lambda} +  \partial_{\mu}\left(\bar f_{\bar B}  +  \bar f_{\bar T} \right)\delta S_{\nu}^{\ \mu \lambda}, \\
\delta e &= \bar e\bar E^{\mu}_{\ B}\delta e^{B}_{\ \mu}, \quad \delta F^{\lambda}_{\ \nu} = \frac{1}{\bar e}\bar e^{A}_{\ \nu}\bar{f}_{\bar T} \partial_{\mu}\left(\bar{e}\delta S_{A}^{\ \mu \lambda} + \delta e \bar{S}_{A}^{\ \mu \lambda} \right) + \left[\frac{1}{\bar e}\bar{e}^A_{\ \nu}\delta f_T + \left( \frac{1}{\bar{e}}\delta e^{A}_{\ \nu} - \frac{\delta e}{\bar{e}^2}\bar{e}^{A}_{\ \nu}\right)\bar f_{\bar T} \right]\partial_{\mu}\left(\bar e \bar S_{A}^{\ \mu \lambda} \right), \\
\delta H^{\lambda}_{\ \nu} &= \bar{S}_{\alpha}^{\ \lambda \mu}\left[\bar f_{\bar T}\delta T^{\alpha}_{\mu \nu}+ \delta f_T \bar{T}^{\alpha}_{\mu \nu} \right] + \bar{f}_{\bar T}\bar T^{\alpha}_{\mu \nu}\delta S_{\alpha}^{\ \lambda \mu},
\end{align}
where
\begin{align} \label{PChrySymbTG}
\delta  \accentset{\circ}{\Gamma}^{\rho}{}_{\mu\nu} &= \frac{1}{2}\bar{g}^{\rho\sigma}\left( \accentset{\circ}{\bar{\nabla}}_{\mu}\delta g_{\sigma\nu} +  \accentset{\circ}{\bar{\nabla}}_{\nu}\delta g_{\mu\sigma} -  \accentset{\circ}{\bar{\nabla}}_{\sigma}\delta g_{\mu\nu}\right) = \bar{g}^{\rho\sigma}\left( \accentset{\circ}{\bar{\nabla}}_{\mu}\tau_{(\sigma\nu)} +  \accentset{\circ}{\bar{\nabla}}_{\nu}\tau_{(\mu\sigma)} -  \accentset{\circ}{\bar{\nabla}}_{\sigma}\tau_{(\mu\nu)}\right) \\
&=\bar{ g}^{\rho \sigma}\left(\tau_{(\sigma \mu),\nu} + \tau_{(\sigma \nu),\mu} -\tau_{(\mu \nu),\sigma} -2\tau_{(\sigma \alpha)} \accentset{\circ}{\bar{\Gamma}}^{\alpha}_{\ \mu \nu} \right),
\end{align}
$\tau_{\mu \nu}$ given by \eqref{pmetricfromtetrtad}, the bar indicates that the corresponding geometrical is object is computed with the background tetrad, and the perturbations for the functional are
\begin{align}
f(T,B) = f(\bar T + \delta T, \bar B + \delta B) = \bar f + \bar f_{\bar T}\delta T + \bar f_{\bar B}\delta B \equiv \bar f +\delta f,
 \end{align}
 where we have defined $\bar f = f(\bar T, \bar B)$ and $\delta f =  \bar f_{\bar T}\delta T + \bar f_{\bar B}\delta B $, and for $f_T$ and $f_B$ we have
 \begin{align}
 f_T(T,B) = \bar f_{\bar{T}} + \delta f_T, \quad f_B(T,B) = \bar{f}_{\bar B} + \delta f_B
 \end{align}
 where 
 \begin{align}
 \delta f_T = \bar f_{\bar T \bar T} \delta T + \bar f_{\bar T \bar B}\delta B \quad \text{and} \quad \delta f_B =  \bar f_{\bar B \bar T} \delta T + \bar f_{\bar B \bar B}\delta B.
 \end{align}
 For the tensor perturbation  given by eq. \eqref{ptetradtensor}, we obtain
 
\begin{eqnarray}
&& \delta T^{i}_{\ 0 j} = \frac{1}{2}h_{ij}', \quad \delta T^{i}_{\ jk} = \frac{1}{2}(\partial_j h_{ik} - \partial_{k}h_{ij}), \delta S_{0}^{\ 0i}=0, \quad \delta S_{i}^{\ 0j} = \frac{1}{4a^{2}}h_{ij}', 
\\
&& \delta S_{i}^{\ jk} = -\frac{1}{4a^2}(\partial_{j}h_{ik} - \partial_k h_{ij}), \quad \delta T = \delta B = 0,
\end{eqnarray}
and then using the aforementioned perturbed quantities, the corresponding perturbed geometrical part of the field equations is
\begin{align}
\delta W^{i}_{\ j} = \frac{f_T}{2a^2} \delta ^{ik}\left(h''_{kj} + [2 + \nu]\mathcal{H}h'_{kj} - \nabla^2 h_{kj} \right), \label{perturbedFieldEq}
\end{align}
giving us the GW propagation equation in Fourier space as shown in eq. \eqref{GWTG}.

\bibliographystyle{model1-num-names}
\bibliography{refs2}

\begin{thebibliography}{81}
\expandafter\ifx\csname natexlab\endcsname\relax\def\natexlab#1{#1}\fi
\providecommand{\url}[1]{\texttt{#1}}
\providecommand{\href}[2]{#2}
\providecommand{\path}[1]{#1}
\providecommand{\DOIprefix}{doi:}
\providecommand{\ArXivprefix}{arXiv:}
\providecommand{\URLprefix}{URL: }
\providecommand{\Pubmedprefix}{pmid:}
\providecommand{\doi}[1]{\href{http://dx.doi.org/#1}{\path{#1}}}
\providecommand{\Pubmed}[1]{\href{pmid:#1}{\path{#1}}}
\providecommand{\bibinfo}[2]{#2}
\ifx\xfnm\relax \def\xfnm[#1]{\unskip,\space#1}\fi
\bibitem[{Clifton et~al.(2012)Clifton, Ferreira, Padilla, and
  Skordis}]{Clifton:2011jh}
\bibinfo{author}{T.~Clifton}, \bibinfo{author}{P.~G. Ferreira},
  \bibinfo{author}{A.~Padilla}, \bibinfo{author}{C.~Skordis},
\newblock \bibinfo{title}{{Modified Gravity and Cosmology}},
\newblock \bibinfo{journal}{Phys. Rept.} \bibinfo{volume}{513}
  (\bibinfo{year}{2012}) \bibinfo{pages}{1--189}.
\bibitem[{Aghanim et~al.(2020)}]{Planck2018}
\bibinfo{author}{N.~Aghanim}, et~al. (\bibinfo{collaboration}{Planck}),
\newblock \bibinfo{title}{{Planck 2018 results. VI. Cosmological parameters}},
\newblock \bibinfo{journal}{Astron. Astrophys.} \bibinfo{volume}{641}
  (\bibinfo{year}{2020}) \bibinfo{pages}{A6}.
\bibitem[{Tr\"{o}ster et~al.(2020)}]{Troster2020}
\bibinfo{author}{T.~Tr\"{o}ster}, et~al.,
\newblock \bibinfo{title}{Cosmology from large-scale structure},
\newblock \bibinfo{journal}{Astronomy $\&$ Astrophysics} \bibinfo{volume}{633}
  (\bibinfo{year}{2020}) \bibinfo{pages}{L10}.
\bibitem[{Dodelson and Schmidt(2021)}]{Dodelson2021-mi}
\bibinfo{author}{S.~Dodelson}, \bibinfo{author}{F.~Schmidt},
  \bibinfo{title}{Modern Cosmology}, \bibinfo{publisher}{Elsevier},
  \bibinfo{year}{2021}.
\bibitem[{Sahni(2004)}]{Sahni2004}
\bibinfo{author}{V.~Sahni},
\newblock \bibinfo{title}{5 dark matter and dark energy},
\newblock in: \bibinfo{booktitle}{The Physics of the Early Universe},
  \bibinfo{publisher}{Springer Berlin Heidelberg}, \bibinfo{year}{2004}, pp.
  \bibinfo{pages}{141--179}. \URLprefix
  \url{https://doi.org/10.1007/978-3-540-31535-3_5}.
  \DOIprefix\doi{10.1007/978-3-540-31535-3_5}.
\bibitem[{Riess et~al.(1998)}]{Riess1998}
\bibinfo{author}{A.~G. Riess}, et~al.,
\newblock \bibinfo{title}{Observational evidence from supernovae for an
  accelerating universe and a cosmological constant},
\newblock \bibinfo{journal}{The Astronomical Journal} \bibinfo{volume}{116}
  (\bibinfo{year}{1998}) \bibinfo{pages}{1009--1038}.
\bibitem[{Bull et~al.(2016)}]{Bull2016}
\bibinfo{author}{P.~Bull}, et~al.,
\newblock \bibinfo{title}{{Beyond {$\Lambda$}CDM: Problems, solutions, and the
  road ahead}},
\newblock \bibinfo{journal}{Physics of the Dark Universe} \bibinfo{volume}{12}
  (\bibinfo{year}{2016}) \bibinfo{pages}{{56--99}}.
\bibitem[{{Fazlollah Hajkarim and J\"{u}rgen
  Schaffner-Bielich}(2020)}]{Hajkarim2020}
\bibinfo{author}{{Fazlollah Hajkarim and J\"{u}rgen Schaffner-Bielich}},
\newblock \bibinfo{title}{Thermal history of the early universe and primordial
  gravitational waves from induced scalar perturbations},
\newblock \bibinfo{journal}{Physical Review D} \bibinfo{volume}{101}
  (\bibinfo{year}{2020}).
\bibitem[{Perivolaropoulos and Skara(2022)}]{Perivolaropoulos2022}
\bibinfo{author}{L.~Perivolaropoulos}, \bibinfo{author}{F.~Skara},
\newblock \bibinfo{title}{Challenges for {$\Lambda$}{CDM}: An update},
\newblock \bibinfo{journal}{New Astronomy Reviews} \bibinfo{volume}{95}
  (\bibinfo{year}{2022}) \bibinfo{pages}{101659}.
\bibitem[{Martin(2012)}]{Martin2012}
\bibinfo{author}{J.~Martin},
\newblock \bibinfo{title}{Everything you always wanted to know about the
  cosmological constant problem (but were afraid to ask)},
\newblock \bibinfo{journal}{Comptes Rendus Physique} \bibinfo{volume}{13}
  (\bibinfo{year}{2012}) \bibinfo{pages}{566--665}.
\bibitem[{Velten et~al.(2014)Velten, vom Marttens, and Zimdahl}]{Velten2014}
\bibinfo{author}{H.~E.~S. Velten}, \bibinfo{author}{R.~F. vom Marttens},
  \bibinfo{author}{W.~Zimdahl},
\newblock \bibinfo{title}{Aspects of the cosmological
  {\textquotedblleft}coincidence problem{\textquotedblright}},
\newblock \bibinfo{journal}{The European Physical Journal C}
  \bibinfo{volume}{74} (\bibinfo{year}{2014}).
\bibitem[{Weinberg(1989)}]{Weinberg1989}
\bibinfo{author}{S.~Weinberg},
\newblock \bibinfo{title}{The cosmological constant problem},
\newblock \bibinfo{journal}{Reviews of Modern Physics} \bibinfo{volume}{61}
  (\bibinfo{year}{1989}) \bibinfo{pages}{1--23}.
\bibitem[{Valentino et~al.(2021)}]{DiValentino2021}
\bibinfo{author}{E.~D. Valentino}, et~al.,
\newblock \bibinfo{title}{Snowmass2021 - letter of interest cosmology
  intertwined {II}: The hubble constant tension},
\newblock \bibinfo{journal}{Astroparticle Physics} \bibinfo{volume}{131}
  (\bibinfo{year}{2021}) \bibinfo{pages}{102605}.
\bibitem[{Abdalla et~al.(2022)}]{Abdalla2022}
\bibinfo{author}{E.~Abdalla}, et~al.,
\newblock \bibinfo{title}{Cosmology intertwined: A review of the particle
  physics, astrophysics, and cosmology associated with the cosmological
  tensions and anomalies},
\newblock \bibinfo{journal}{Journal of High Energy Astrophysics}
  \bibinfo{volume}{34} (\bibinfo{year}{2022}) \bibinfo{pages}{49--211}.
\bibitem[{Ti{\'{a}}n(2020)}]{Tin2020}
\bibinfo{author}{S.~X. Ti{\'{a}}n},
\newblock \bibinfo{title}{Cosmological consequences of a scalar field with
  oscillating equation of state: A possible solution to the fine-tuning and
  coincidence problems},
\newblock \bibinfo{journal}{Physical Review D} \bibinfo{volume}{101}
  (\bibinfo{year}{2020}).
\bibitem[{Chimento et~al.(2003)Chimento, Jakubi, Pav{\'{o}}n, and
  Zimdahl}]{Chimento2003}
\bibinfo{author}{L.~P. Chimento}, \bibinfo{author}{A.~S. Jakubi},
  \bibinfo{author}{D.~Pav{\'{o}}n}, \bibinfo{author}{W.~Zimdahl},
\newblock \bibinfo{title}{Interacting quintessence solution to the coincidence
  problem},
\newblock \bibinfo{journal}{Physical Review D} \bibinfo{volume}{67}
  (\bibinfo{year}{2003}).
\bibitem[{Kang et~al.(2020)Kang, Zhang, Jun, and Zong}]{Kang2020}
\bibinfo{author}{G.-Z. Kang}, \bibinfo{author}{D.-S. Zhang},
  \bibinfo{author}{L.~Jun}, \bibinfo{author}{H.-S. Zong},
\newblock \bibinfo{title}{Fine tuning problem of the cosmological constant in a
  generalized randall-sundrum model},
\newblock \bibinfo{journal}{Chinese Physics C} \bibinfo{volume}{44}
  (\bibinfo{year}{2020}) \bibinfo{pages}{125102}.
\bibitem[{Feng and Li(2014)}]{Feng2014}
\bibinfo{author}{C.-J. Feng}, \bibinfo{author}{X.-Z. Li},
\newblock \bibinfo{title}{Towards a realistic solution of the cosmological
  constant fine-tuning problem by higgs inflation},
\newblock \bibinfo{journal}{Physical Review D} \bibinfo{volume}{90}
  (\bibinfo{year}{2014}).
\bibitem[{Bisabr(2010)}]{Bisabr2010}
\bibinfo{author}{Y.~Bisabr},
\newblock \bibinfo{title}{Coincidence problem in {$f(R)$} gravity models},
\newblock \bibinfo{journal}{Physical Review D} \bibinfo{volume}{82}
  (\bibinfo{year}{2010}).
\bibitem[{Rudra(2015)}]{Rudra2015}
\bibinfo{author}{P.~Rudra},
\newblock \bibinfo{title}{Towards a possible solution for the coincidence
  problem: f(g) gravity as background},
\newblock \bibinfo{journal}{International Journal of Modern Physics D}
  \bibinfo{volume}{24} (\bibinfo{year}{2015}) \bibinfo{pages}{1550013}.
\bibitem[{Krishnan et~al.(2022)Krishnan, Mohayaee, Colg\'ain, Sheikh-Jabbari,
  and Yin}]{Krishnan:2021jmh}
\bibinfo{author}{C.~Krishnan}, \bibinfo{author}{R.~Mohayaee},
  \bibinfo{author}{E.~O. Colg\'ain}, \bibinfo{author}{M.~M. Sheikh-Jabbari},
  \bibinfo{author}{L.~Yin},
\newblock \bibinfo{title}{{Hints of FLRW breakdown from supernovae}},
\newblock \bibinfo{journal}{Phys. Rev. D} \bibinfo{volume}{105}
  (\bibinfo{year}{2022}) \bibinfo{pages}{063514}.
\bibitem[{Krishnan et~al.(2021)Krishnan, Mohayaee, Colg\'ain, Sheikh-Jabbari,
  and Yin}]{Krishnan:2021dyb}
\bibinfo{author}{C.~Krishnan}, \bibinfo{author}{R.~Mohayaee},
  \bibinfo{author}{E.~O. Colg\'ain}, \bibinfo{author}{M.~M. Sheikh-Jabbari},
  \bibinfo{author}{L.~Yin},
\newblock \bibinfo{title}{{Does Hubble tension signal a breakdown in FLRW
  cosmology?}},
\newblock \bibinfo{journal}{Class. Quant. Grav.} \bibinfo{volume}{38}
  (\bibinfo{year}{2021}) \bibinfo{pages}{184001}.
\bibitem[{Poulin et~al.(2023)Poulin, Smith, and Karwal}]{Poulin:2023lkg}
\bibinfo{author}{V.~Poulin}, \bibinfo{author}{T.~L. Smith},
  \bibinfo{author}{T.~Karwal},
\newblock \bibinfo{title}{{The Ups and Downs of Early Dark Energy solutions to
  the Hubble tension: a review of models, hints and constraints circa 2023}}
  (\bibinfo{year}{2023}).
\bibitem[{Di~Valentino and Melchiorri(2022)}]{DiValentino:2021imh}
\bibinfo{author}{E.~Di~Valentino}, \bibinfo{author}{A.~Melchiorri},
\newblock \bibinfo{title}{{Neutrino Mass Bounds in the Era of Tension
  Cosmology}},
\newblock \bibinfo{journal}{Astrophys. J. Lett.} \bibinfo{volume}{931}
  (\bibinfo{year}{2022}) \bibinfo{pages}{L18}.
\bibitem[{Di~Valentino et~al.(2022)Di~Valentino, Gariazzo, Giunti, Mena, Pan,
  and Yang}]{DiValentino:2021rjj}
\bibinfo{author}{E.~Di~Valentino}, \bibinfo{author}{S.~Gariazzo},
  \bibinfo{author}{C.~Giunti}, \bibinfo{author}{O.~Mena},
  \bibinfo{author}{S.~Pan}, \bibinfo{author}{W.~Yang},
\newblock \bibinfo{title}{{Minimal dark energy: Key to sterile neutrino and
  Hubble constant tensions?}},
\newblock \bibinfo{journal}{Phys. Rev. D} \bibinfo{volume}{105}
  (\bibinfo{year}{2022}) \bibinfo{pages}{103511}.
\bibitem[{Aldrovandi and Pereira(2013)}]{Aldrovandi:2013wha}
\bibinfo{author}{R.~Aldrovandi}, \bibinfo{author}{J.~G. Pereira},
  \bibinfo{title}{{Teleparallel Gravity}}, volume \bibinfo{volume}{173},
  \bibinfo{publisher}{Springer}, \bibinfo{address}{Dordrecht},
  \bibinfo{year}{2013}. \DOIprefix\doi{10.1007/978-94-007-5143-9}.
\bibitem[{Bahamonde et~al.(2021)Bahamonde, Dialektopoulos, Escamilla-Rivera,
  Farrugia, Gakis, Hendry, Hohmann, Said, Mifsud, and
  Di~Valentino}]{Bahamonde:2021gfp}
\bibinfo{author}{S.~Bahamonde}, \bibinfo{author}{K.~F. Dialektopoulos},
  \bibinfo{author}{C.~Escamilla-Rivera}, \bibinfo{author}{G.~Farrugia},
  \bibinfo{author}{V.~Gakis}, \bibinfo{author}{M.~Hendry},
  \bibinfo{author}{M.~Hohmann}, \bibinfo{author}{J.~L. Said},
  \bibinfo{author}{J.~Mifsud}, \bibinfo{author}{E.~Di~Valentino},
\newblock \bibinfo{title}{{Teleparallel Gravity: From Theory to Cosmology}}
  (\bibinfo{year}{2021}).
\bibitem[{Krssak et~al.(2019)Krssak, van~den Hoogen, Pereira, Böhmer, and
  Coley}]{Krssak:2018ywd}
\bibinfo{author}{M.~Krssak}, \bibinfo{author}{R.~van~den Hoogen},
  \bibinfo{author}{J.~Pereira}, \bibinfo{author}{C.~Böhmer},
  \bibinfo{author}{A.~Coley},
\newblock \bibinfo{title}{{Teleparallel theories of gravity: illuminating a
  fully invariant approach}},
\newblock \bibinfo{journal}{Class. Quant. Grav.} \bibinfo{volume}{36}
  (\bibinfo{year}{2019}) \bibinfo{pages}{183001}.
\bibitem[{Cai et~al.(2016)Cai, Capozziello, De~Laurentis, and
  Saridakis}]{Cai:2015emx}
\bibinfo{author}{Y.-F. Cai}, \bibinfo{author}{S.~Capozziello},
  \bibinfo{author}{M.~De~Laurentis}, \bibinfo{author}{E.~N. Saridakis},
\newblock \bibinfo{title}{{$f(T)$ teleparallel gravity and cosmology}},
\newblock \bibinfo{journal}{Rept. Prog. Phys.} \bibinfo{volume}{79}
  (\bibinfo{year}{2016}) \bibinfo{pages}{106901}.
\bibitem[{Kadam et~al.(2022)Kadam, Said, and Mishra}]{Kadam2022}
\bibinfo{author}{S.~A. Kadam}, \bibinfo{author}{J.~L. Said},
  \bibinfo{author}{B.~Mishra},
\newblock \bibinfo{title}{Accelerating cosmological models in {$f(T, B)$}
  gravitational theory}  (\bibinfo{year}{2022}).
\bibitem[{Franco et~al.(2020)Franco, Escamilla-Rivera, and
  Levi~Said}]{Franco2020}
\bibinfo{author}{G.~A.~R. Franco}, \bibinfo{author}{C.~Escamilla-Rivera},
  \bibinfo{author}{J.~Levi~Said},
\newblock \bibinfo{title}{{Stability analysis for cosmological models in $f(T,
  B)$ gravity}},
\newblock \bibinfo{journal}{Eur. Phys. J. C} \bibinfo{volume}{80}
  (\bibinfo{year}{2020}) \bibinfo{pages}{677}.
\bibitem[{Mirza and Oboudiat(2017)}]{Mirza2017}
\bibinfo{author}{B.~Mirza}, \bibinfo{author}{F.~Oboudiat},
\newblock \bibinfo{title}{Constraining f(t) gravity by dynamical system
  analysis},
\newblock \bibinfo{journal}{Journal of Cosmology and Astroparticle Physics}
  \bibinfo{volume}{2017} (\bibinfo{year}{2017}) \bibinfo{pages}{011--011}.
\bibitem[{Rave-Franco et~al.(2021)Rave-Franco, Escamilla-Rivera, and
  Said}]{RaveFranco2021}
\bibinfo{author}{G.~A. Rave-Franco}, \bibinfo{author}{C.~Escamilla-Rivera},
  \bibinfo{author}{J.~L. Said},
\newblock \bibinfo{title}{Dynamical complexity of the teleparallel gravity
  cosmology},
\newblock \bibinfo{journal}{Physical Review D} \bibinfo{volume}{103}
  (\bibinfo{year}{2021}).
\bibitem[{Briffa et~al.(2022)Briffa, Escamilla-Rivera, Said, Mifsud, and
  Pullicino}]{Briffa2022}
\bibinfo{author}{R.~Briffa}, \bibinfo{author}{C.~Escamilla-Rivera},
  \bibinfo{author}{J.~L. Said}, \bibinfo{author}{J.~Mifsud},
  \bibinfo{author}{N.~L. Pullicino},
\newblock \bibinfo{title}{Impact of {$H_0$} priors on f(t) late time
  cosmology},
\newblock \bibinfo{journal}{The European Physical Journal Plus}
  \bibinfo{volume}{137} (\bibinfo{year}{2022}).
\bibitem[{Nunes(2018)}]{Nunes2018}
\bibinfo{author}{R.~C. Nunes},
\newblock \bibinfo{title}{Structure formation in {$f(T)$} gravity and a
  solution for {$H_0$} tension},
\newblock \bibinfo{journal}{Journal of Cosmology and Astroparticle Physics}
  \bibinfo{volume}{2018} (\bibinfo{year}{2018}) \bibinfo{pages}{052--052}.
\bibitem[{N{\'{a}}jera et~al.(2022)N{\'{a}}jera, Aguilar, Rave-Franco,
  Escamilla-Rivera, and Sussman}]{Njera2022}
\bibinfo{author}{S.~N{\'{a}}jera}, \bibinfo{author}{A.~Aguilar},
  \bibinfo{author}{G.~A. Rave-Franco}, \bibinfo{author}{C.~Escamilla-Rivera},
  \bibinfo{author}{R.~A. Sussman},
\newblock \bibinfo{title}{Inhomogeneous solutions in {$f(T, B)$} gravity},
\newblock \bibinfo{journal}{International Journal of Geometric Methods in
  Modern Physics} \bibinfo{volume}{19} (\bibinfo{year}{2022}).
\bibitem[{Bourakadi et~al.(2022)Bourakadi, Asfour, Sakhi, Bennai, and
  Ouali}]{ElBourakadi2022}
\bibinfo{author}{K.~E. Bourakadi}, \bibinfo{author}{B.~Asfour},
  \bibinfo{author}{Z.~Sakhi}, \bibinfo{author}{M.~Bennai},
  \bibinfo{author}{T.~Ouali},
\newblock \bibinfo{title}{Primordial black holes and gravitational waves in
  teleparallel gravity},
\newblock \bibinfo{journal}{The European Physical Journal C}
  \bibinfo{volume}{82} (\bibinfo{year}{2022}).
\bibitem[{Sahlu et~al.(2020)Sahlu, Ntahompagaze, Abebe, and Mota}]{Sahlu2020}
\bibinfo{author}{S.~Sahlu}, \bibinfo{author}{J.~Ntahompagaze},
  \bibinfo{author}{A.~Abebe}, \bibinfo{author}{D.~F. Mota},
\newblock \bibinfo{title}{Inflationary constraints in teleparallel gravity
  theory},
\newblock \bibinfo{journal}{International Journal of Geometric Methods in
  Modern Physics} \bibinfo{volume}{18} (\bibinfo{year}{2020})
  \bibinfo{pages}{2150027}.
\bibitem[{Raatikainen and R\"{a}s\"{a}nen(2019)}]{Raatikainen2019}
\bibinfo{author}{S.~Raatikainen}, \bibinfo{author}{S.~R\"{a}s\"{a}nen},
\newblock \bibinfo{title}{Higgs inflation and teleparallel gravity},
\newblock \bibinfo{journal}{Journal of Cosmology and Astroparticle Physics}
  \bibinfo{volume}{2019} (\bibinfo{year}{2019}) \bibinfo{pages}{021--021}.
\bibitem[{Chakrabortty et~al.(2021)Chakrabortty, Sk, Sanyal, and
  Sanyal}]{Chakrabortty2021}
\bibinfo{author}{M.~Chakrabortty}, \bibinfo{author}{N.~Sk},
  \bibinfo{author}{S.~Sanyal}, \bibinfo{author}{A.~K. Sanyal},
\newblock \bibinfo{title}{Inflation with f(t) teleparallel gravity},
\newblock \bibinfo{journal}{The European Physical Journal Plus}
  \bibinfo{volume}{136} (\bibinfo{year}{2021}).
\bibitem[{Ricciardone(2017)}]{Ricciardone2017}
\bibinfo{author}{A.~Ricciardone},
\newblock \bibinfo{title}{Primordial gravitational waves with {LISA}},
\newblock \bibinfo{journal}{Journal of Physics: Conference Series}
  \bibinfo{volume}{840} (\bibinfo{year}{2017}) \bibinfo{pages}{012030}.
\bibitem[{Hohmann et~al.(2019)Hohmann, J\"{a}rv, Kr{\v{s}}{\v{s}}{\'{a}}k, and
  Pfeifer}]{Hohmann2019}
\bibinfo{author}{M.~Hohmann}, \bibinfo{author}{L.~J\"{a}rv},
  \bibinfo{author}{M.~Kr{\v{s}}{\v{s}}{\'{a}}k}, \bibinfo{author}{C.~Pfeifer},
\newblock \bibinfo{title}{Modified teleparallel theories of gravity in
  symmetric spacetimes},
\newblock \bibinfo{journal}{Physical Review D} \bibinfo{volume}{100}
  (\bibinfo{year}{2019}).
\bibitem[{Socolovsky(2012{\natexlab{a}})}]{Socolovsky2012a}
\bibinfo{author}{M.~Socolovsky},
\newblock \bibinfo{title}{Fiber bundles, connections, general relativity, and
  the einstein-cartan theory {\textendash} part i},
\newblock \bibinfo{journal}{Advances in Applied Clifford Algebras}
  \bibinfo{volume}{22} (\bibinfo{year}{2012}{\natexlab{a}})
  \bibinfo{pages}{837--872}.
\bibitem[{Socolovsky(2012{\natexlab{b}})}]{Socolovsky2012b}
\bibinfo{author}{M.~Socolovsky},
\newblock \bibinfo{title}{Fiber bundles, connections, general relativity, and
  the einstein-cartan theory {\textendash} part {II}},
\newblock \bibinfo{journal}{Advances in Applied Clifford Algebras}
  \bibinfo{volume}{22} (\bibinfo{year}{2012}{\natexlab{b}})
  \bibinfo{pages}{873--909}.
\bibitem[{Blau(2011)}]{Blau2011}
\bibinfo{author}{M.~Blau}, \bibinfo{title}{{L}ecture {N}otes on {G}eneral
  {R}elativity}, \bibinfo{type}{Technical Report}, Albert Einstein Center for
  Fundamental Physics, University of Bern, \bibinfo{year}{2011}. \URLprefix
  \url{http://www.blau.itp.unibe.ch/newlecturesGR.pdf}.
\bibitem[{Golovnev et~al.(2017)Golovnev, Koivisto, and Sandstad}]{Golovnev2017}
\bibinfo{author}{A.~Golovnev}, \bibinfo{author}{T.~Koivisto},
  \bibinfo{author}{M.~Sandstad},
\newblock \bibinfo{title}{On the covariance of teleparallel gravity theories},
\newblock \bibinfo{journal}{Classical and Quantum Gravity} \bibinfo{volume}{34}
  (\bibinfo{year}{2017}) \bibinfo{pages}{145013}.
\bibitem[{Carroll(2004)}]{Carroll2004}
\bibinfo{author}{S.~Carroll}, \bibinfo{title}{Spacetime and Geometry: An
  Introduction to General Relativity}, \bibinfo{publisher}{Addison Wesley},
  \bibinfo{year}{2004}.
\bibitem[{Sotiriou and Faraoni(2010)}]{Sotiriou:2008rp}
\bibinfo{author}{T.~P. Sotiriou}, \bibinfo{author}{V.~Faraoni},
\newblock \bibinfo{title}{{f(R) Theories Of Gravity}},
\newblock \bibinfo{journal}{Rev. Mod. Phys.} \bibinfo{volume}{82}
  (\bibinfo{year}{2010}) \bibinfo{pages}{451--497}.
\bibitem[{Faraoni(2008)}]{Faraoni:2008mf}
\bibinfo{author}{V.~Faraoni},
\newblock \bibinfo{title}{{f(R) gravity: Successes and challenges}},
\newblock in: \bibinfo{booktitle}{{18th SIGRAV Conference}},
  \bibinfo{year}{2008}. \href{http://arxiv.org/abs/0810.2602}{\tt
  arXiv:0810.2602}.
\bibitem[{Capozziello et~al.(2018)Capozziello, Capriolo, and
  Transirico}]{Capozziello:2018qcp}
\bibinfo{author}{S.~Capozziello}, \bibinfo{author}{M.~Capriolo},
  \bibinfo{author}{M.~Transirico},
\newblock \bibinfo{title}{{The gravitational energy-momentum pseudotensor: the
  cases of $f(R)$ and $f(T)$ gravity}},
\newblock \bibinfo{journal}{Int. J. Geom. Meth. Mod. Phys.}
  \bibinfo{volume}{15} (\bibinfo{year}{2018}) \bibinfo{pages}{1850164}.
\bibitem[{Farrugia et~al.(2020)Farrugia, Levi~Said, and
  Finch}]{Farrugia:2020fcu}
\bibinfo{author}{G.~Farrugia}, \bibinfo{author}{J.~Levi~Said},
  \bibinfo{author}{A.~Finch},
\newblock \bibinfo{title}{{Gravitoelectromagnetism, Solar System Test and
  Weak-Field Solutions in $f(T,B)$ Gravity with Observational Constraints}},
\newblock \bibinfo{journal}{Universe} \bibinfo{volume}{6}
  (\bibinfo{year}{2020}) \bibinfo{pages}{34}.
\bibitem[{Farrugia et~al.(2018)Farrugia, Levi~Said, Gakis, and
  Saridakis}]{Farrugia:2018gyz}
\bibinfo{author}{G.~Farrugia}, \bibinfo{author}{J.~Levi~Said},
  \bibinfo{author}{V.~Gakis}, \bibinfo{author}{E.~N. Saridakis},
\newblock \bibinfo{title}{{Gravitational Waves in Modified Teleparallel
  Theories}},
\newblock \bibinfo{journal}{Phys. Rev. D} \bibinfo{volume}{97}
  (\bibinfo{year}{2018}) \bibinfo{pages}{124064}.
\bibitem[{Bahamonde et~al.(2015)Bahamonde, Böhmer, and
  Wright}]{Bahamonde:2015zma}
\bibinfo{author}{S.~Bahamonde}, \bibinfo{author}{C.~G. Böhmer},
  \bibinfo{author}{M.~Wright},
\newblock \bibinfo{title}{{Modified teleparallel theories of gravity}},
\newblock \bibinfo{journal}{Phys. Rev.} \bibinfo{volume}{D92}
  (\bibinfo{year}{2015}) \bibinfo{pages}{104042}.
\bibitem[{Wald(1984)}]{Wald}
\bibinfo{author}{R.~M. Wald}, \bibinfo{title}{General Relativity},
  \bibinfo{publisher}{University of Chicago Pr.}, \bibinfo{year}{1984}.
\bibitem[{Bahamonde et~al.(2021)Bahamonde, Gakis, Kiorpelidi, Koivisto, Said,
  and Saridakis}]{Bahamonde2021}
\bibinfo{author}{S.~Bahamonde}, \bibinfo{author}{V.~Gakis},
  \bibinfo{author}{S.~Kiorpelidi}, \bibinfo{author}{T.~Koivisto},
  \bibinfo{author}{J.~L. Said}, \bibinfo{author}{E.~N. Saridakis},
\newblock \bibinfo{title}{Cosmological perturbations in modified teleparallel
  gravity models: boundary term extension},
\newblock \bibinfo{journal}{The European Physical Journal C}
  \bibinfo{volume}{81} (\bibinfo{year}{2021}).
\bibitem[{Piattella(2018)}]{Piattella2018}
\bibinfo{author}{O.~Piattella}, \bibinfo{title}{Lecture Notes in Cosmology},
  \bibinfo{publisher}{Springer International Publishing}, \bibinfo{year}{2018}.
  \URLprefix \url{https://doi.org/10.1007/978-3-319-95570-4}.
  \DOIprefix\doi{10.1007/978-3-319-95570-4}.
\bibitem[{Hohmann(2021)}]{Hohmann2021}
\bibinfo{author}{M.~Hohmann},
\newblock \bibinfo{title}{General cosmological perturbations in teleparallel
  gravity},
\newblock \bibinfo{journal}{The European Physical Journal Plus}
  \bibinfo{volume}{136} (\bibinfo{year}{2021}).
\bibitem[{Chen et~al.(2011)Chen, Dent, Dutta, and Saridakis}]{Chen:2010va}
\bibinfo{author}{S.-H. Chen}, \bibinfo{author}{J.~B. Dent},
  \bibinfo{author}{S.~Dutta}, \bibinfo{author}{E.~N. Saridakis},
\newblock \bibinfo{title}{{Cosmological perturbations in f(T) gravity}},
\newblock \bibinfo{journal}{Phys. Rev.} \bibinfo{volume}{D83}
  (\bibinfo{year}{2011}) \bibinfo{pages}{023508}.
\bibitem[{Bahamonde et~al.(2023)Bahamonde, Dialektopoulos, Hohmann, Levi~Said,
  Pfeifer, and Saridakis}]{Bahamonde:2022ohm}
\bibinfo{author}{S.~Bahamonde}, \bibinfo{author}{K.~F. Dialektopoulos},
  \bibinfo{author}{M.~Hohmann}, \bibinfo{author}{J.~Levi~Said},
  \bibinfo{author}{C.~Pfeifer}, \bibinfo{author}{E.~N. Saridakis},
\newblock \bibinfo{title}{{Perturbations in non-flat cosmology for f(T)
  gravity}},
\newblock \bibinfo{journal}{Eur. Phys. J. C} \bibinfo{volume}{83}
  (\bibinfo{year}{2023}) \bibinfo{pages}{193}.
\bibitem[{Nunes et~al.(2018)Nunes, Pan, and Saridakis}]{Nunes:2018evm}
\bibinfo{author}{R.~C. Nunes}, \bibinfo{author}{S.~Pan}, \bibinfo{author}{E.~N.
  Saridakis},
\newblock \bibinfo{title}{{New observational constraints on $f(T)$ gravity
  through gravitational-wave astronomy}},
\newblock \bibinfo{journal}{Phys. Rev. D} \bibinfo{volume}{98}
  (\bibinfo{year}{2018}) \bibinfo{pages}{104055}.
\bibitem[{Ryden(2017)}]{Ryden2017}
\bibinfo{author}{B.~Ryden}, \bibinfo{title}{Introduction to Cosmology},
  \bibinfo{publisher}{Cambridge University Press}, \bibinfo{year}{2017}.
\bibitem[{Klose et~al.(2022)Klose, Laine, and Procacci}]{Klose2022b}
\bibinfo{author}{P.~Klose}, \bibinfo{author}{M.~Laine},
  \bibinfo{author}{S.~Procacci},
\newblock \bibinfo{title}{Gravitational wave background from non-abelian
  reheating after axion-like inflation},
\newblock \bibinfo{journal}{Journal of Cosmology and Astroparticle Physics}
  \bibinfo{volume}{2022} (\bibinfo{year}{2022}) \bibinfo{pages}{021}.
\bibitem[{Kundu(2012)}]{Kundu2012}
\bibinfo{author}{S.~Kundu},
\newblock \bibinfo{title}{Inflation with general initial conditions for scalar
  perturbations},
\newblock \bibinfo{journal}{Journal of Cosmology and Astroparticle Physics}
  \bibinfo{volume}{2012} (\bibinfo{year}{2012}) \bibinfo{pages}{005--005}.
\bibitem[{Weinberg(2008)}]{weinberg2008cosmology}
\bibinfo{author}{S.~Weinberg}, \bibinfo{title}{Cosmology}, Cosmology,
  \bibinfo{publisher}{OUP Oxford}, \bibinfo{year}{2008}. \URLprefix
  \url{https://books.google.com.mt/books?id=nqQZdg020fsC}.
\bibitem[{Baumann(2022)}]{Baumann2022-sw}
\bibinfo{author}{D.~Baumann}, \bibinfo{title}{Cosmology},
  \bibinfo{publisher}{Cambridge University Press}, \bibinfo{address}{Cambridge,
  England}, \bibinfo{year}{2022}.
\bibitem[{Boyle and Steinhardt(2008)}]{Boyle2008}
\bibinfo{author}{L.~A. Boyle}, \bibinfo{author}{P.~J. Steinhardt},
\newblock \bibinfo{title}{Probing the early universe with inflationary
  gravitational waves},
\newblock \bibinfo{journal}{Physical Review D} \bibinfo{volume}{77}
  (\bibinfo{year}{2008}).
\bibitem[{Bahamonde and Capozziello(2017)}]{Bahamonde:2016grb}
\bibinfo{author}{S.~Bahamonde}, \bibinfo{author}{S.~Capozziello},
\newblock \bibinfo{title}{{Noether Symmetry Approach in $f(T,B)$ teleparallel
  cosmology}},
\newblock \bibinfo{journal}{Eur. Phys. J.} \bibinfo{volume}{C77}
  (\bibinfo{year}{2017}) \bibinfo{pages}{107}.
\bibitem[{Bahamonde et~al.(2019)Bahamonde, Camci, and
  Capozziello}]{Bahamonde2019}
\bibinfo{author}{S.~Bahamonde}, \bibinfo{author}{U.~Camci},
  \bibinfo{author}{S.~Capozziello},
\newblock \bibinfo{title}{Noether symmetries and boundary terms in extended
  teleparallel gravity cosmology},
\newblock \bibinfo{journal}{Classical and Quantum Gravity} \bibinfo{volume}{36}
  (\bibinfo{year}{2019}) \bibinfo{pages}{065013}.
\bibitem[{Rezazadeh et~al.(2016)Rezazadeh, Abdolmaleki, and
  Karami}]{Rezazadeh2016}
\bibinfo{author}{K.~Rezazadeh}, \bibinfo{author}{A.~Abdolmaleki},
  \bibinfo{author}{K.~Karami},
\newblock \bibinfo{title}{Power-law and intermediate inflationary models in
  f(t)-gravity},
\newblock \bibinfo{journal}{Journal of High Energy Physics}
  \bibinfo{volume}{2016} (\bibinfo{year}{2016}).
\bibitem[{dos Santos et~al.(2022)dos Santos, Gonzalez, and
  Silva}]{dosSantos2022}
\bibinfo{author}{F.~B.~M. dos Santos}, \bibinfo{author}{J.~E. Gonzalez},
  \bibinfo{author}{R.~Silva},
\newblock \bibinfo{title}{Observational constraints on f(t) gravity from
  model-independent data},
\newblock \bibinfo{journal}{The European Physical Journal C}
  \bibinfo{volume}{82} (\bibinfo{year}{2022}).
\bibitem[{Escamilla-Rivera and Levi~Said(2020)}]{Escamilla-Rivera:2019ulu}
\bibinfo{author}{C.~Escamilla-Rivera}, \bibinfo{author}{J.~Levi~Said},
\newblock \bibinfo{title}{{Cosmological viable models in $f(T,B)$ theory as
  solutions to the $H_0$ tension}},
\newblock \bibinfo{journal}{Class. Quant. Grav.} \bibinfo{volume}{37}
  (\bibinfo{year}{2020}) \bibinfo{pages}{165002}.
\bibitem[{Li et~al.(2018)Li, Cai, Cai, and Saridakis}]{Li2018}
\bibinfo{author}{C.~Li}, \bibinfo{author}{Y.~Cai}, \bibinfo{author}{Y.-F. Cai},
  \bibinfo{author}{E.~N. Saridakis},
\newblock \bibinfo{title}{The effective field theory approach of teleparallel
  gravity, $f(t)$ gravity and beyond},
\newblock \bibinfo{journal}{Journal of Cosmology and Astroparticle Physics}
  \bibinfo{volume}{2018} (\bibinfo{year}{2018}) \bibinfo{pages}{001--001}.
\bibitem[{Nesseris et~al.(2013)Nesseris, Basilakos, Saridakis, and
  Perivolaropoulos}]{Nesseris2013}
\bibinfo{author}{S.~Nesseris}, \bibinfo{author}{S.~Basilakos},
  \bibinfo{author}{E.~N. Saridakis}, \bibinfo{author}{L.~Perivolaropoulos},
\newblock \bibinfo{title}{Viable $f(t)$ models are practically
  indistinguishable from lcdm},
\newblock \bibinfo{journal}{Physical Review D} \bibinfo{volume}{88}
  (\bibinfo{year}{2013}).
\bibitem[{Klose et~al.(2022)Klose, Laine, and Procacci}]{Klose2022}
\bibinfo{author}{P.~Klose}, \bibinfo{author}{M.~Laine},
  \bibinfo{author}{S.~Procacci},
\newblock \bibinfo{title}{Gravitational wave background from vacuum and thermal
  fluctuations during axion-like inflation},
\newblock \bibinfo{journal}{Journal of Cosmology and Astroparticle Physics}
  \bibinfo{volume}{2022} (\bibinfo{year}{2022}) \bibinfo{pages}{020}.
\bibitem[{Hogan(1986)}]{Hogan1986}
\bibinfo{author}{C.~J. Hogan},
\newblock \bibinfo{title}{Gravitational radiation from cosmological phase
  transitions},
\newblock \bibinfo{journal}{Monthly Notices of the Royal Astronomical Society}
  \bibinfo{volume}{218} (\bibinfo{year}{1986}) \bibinfo{pages}{629--636}.
\bibitem[{Durrer(2010)}]{Durrer2010}
\bibinfo{author}{R.~Durrer},
\newblock \bibinfo{title}{Gravitational waves from cosmological phase
  transitions},
\newblock \bibinfo{journal}{Journal of Physics: Conference Series}
  \bibinfo{volume}{222} (\bibinfo{year}{2010}) \bibinfo{pages}{012021}.
\bibitem[{Grojean and Servant(2007)}]{Grojean2007}
\bibinfo{author}{C.~Grojean}, \bibinfo{author}{G.~Servant},
\newblock \bibinfo{title}{Gravitational waves from phase transitions at the
  electroweak scale and beyond},
\newblock \bibinfo{journal}{Physical Review D} \bibinfo{volume}{75}
  (\bibinfo{year}{2007}).
\bibitem[{Caprini and Durrer(2006)}]{Caprini2006}
\bibinfo{author}{C.~Caprini}, \bibinfo{author}{R.~Durrer},
\newblock \bibinfo{title}{Gravitational waves from stochastic relativistic
  sources: Primordial turbulence and magnetic fields},
\newblock \bibinfo{journal}{Physical Review D} \bibinfo{volume}{74}
  (\bibinfo{year}{2006}).
\bibitem[{Benetti et~al.(2022)Benetti, Graef, and Vagnozzi}]{Benetti2022}
\bibinfo{author}{M.~Benetti}, \bibinfo{author}{L.~L. Graef},
  \bibinfo{author}{S.~Vagnozzi},
\newblock \bibinfo{title}{Primordial gravitational waves from {NANOGrav}: A
  broken power-law approach},
\newblock \bibinfo{journal}{Physical Review D} \bibinfo{volume}{105}
  (\bibinfo{year}{2022}).
\bibitem[{Papanikolaou et~al.(2023)Papanikolaou, Tzerefos, Basilakos, and
  Saridakis}]{Papanikolaou:2022hkg}
\bibinfo{author}{T.~Papanikolaou}, \bibinfo{author}{C.~Tzerefos},
  \bibinfo{author}{S.~Basilakos}, \bibinfo{author}{E.~N. Saridakis},
\newblock \bibinfo{title}{{No constraints for f(T) gravity from gravitational
  waves induced from primordial black hole fluctuations}},
\newblock \bibinfo{journal}{Eur. Phys. J. C} \bibinfo{volume}{83}
  (\bibinfo{year}{2023}) \bibinfo{pages}{31}.
\bibitem[{Tzerefos et~al.(2023)Tzerefos, Papanikolaou, Saridakis, and
  Basilakos}]{2303.16695}
\bibinfo{author}{C.~Tzerefos}, \bibinfo{author}{T.~Papanikolaou},
  \bibinfo{author}{E.~N. Saridakis}, \bibinfo{author}{S.~Basilakos},
\newblock \bibinfo{title}{Scalar induced gravitational waves in modified
  teleparallel gravity theories}  (\bibinfo{year}{2023}).

\end{thebibliography}

\end{document}